\documentclass[runningheads,a4paper]{llncs}
\usepackage{isabelle,isabellesym}

\usepackage[T1]{fontenc} 
\usepackage[utf8]{inputenc} 
\usepackage[english]{babel} 
\usepackage{lmodern} 

\isabellestyle{it}

\usepackage{pdfsetup}

\usepackage[color]{vdmlisting}

\begin{document}

\title{VDM recursive functions in Isabelle/HOL}
\author{Leo Freitas\inst{1} \and Peter Gorm Larsen\inst{2}}

\authorrunning{ }

\institute{
School of Computing, Newcastle University, \\
\email{leo.freitas@newcastle.ac.uk}
\and 
DIGIT, Aarhus University, Department of Engineering, \\
\email{pgl@ece.au.dk}
}
			
\maketitle
\setcounter{footnote}{0} 
\begin{abstract}
For recursive functions general principles of induction needs to be applied. Instead of verifying them directly using the Vienna Development Method Specification Language (VDM-SL), we suggest a translation to Isabelle/HOL. In this paper, the challenges of such a translation for recursive functions are presented. This is an extension of an existing translation and a VDM mathematical toolbox in Isabelle/HOL enabling support for recursive functions.
\end{abstract}



\parindent 0pt\parskip 0.5ex


%
\begin{isabellebody}%
\setisabellecontext{RecursiveVDM}%
\isadelimtheory
\endisadelimtheory
\isatagtheory
\endisatagtheory
{\isafoldtheory}%
\isadelimtheory
\endisadelimtheory
\isadelimdocument
\endisadelimdocument
\isatagdocument
\isamarkupsection{Introduction\label{sec:Intro}%
}
\isamarkuptrue%
\endisatagdocument
{\isafolddocument}%
\isadelimdocument
\endisadelimdocument
\begin{isamarkuptext}%
This paper describes a translation strategy for a variety of recursive
definitions from VDM to Isabelle/HOL. The strategy takes into account the
differences in how termination and well-foundedness are represented in both
formalisms.

Beyond overcoming technical practicalities, which we discuss, a major
objective is to create translation strategy templates. These templates must
cover a wide variety of VDM recursive definitions, as well as having their
proof obligations being highly automated. The result is an extension to the
VDM to Isabelle/HOL translation strategy and implementation as a plugin to
VDMJ~\cite{Battle09} and extension to VDM-VSCode \cite{AdvancedVSCodePaper}.

Isabelle uses literate programming, where formal specification, proofs and
documentation are all within the same environment. We omit proof scripts
below; the full VDM and Isabelle sources and proofs can be found at the VDM
toolkit repository at
\isatt{.{\kern0pt}/{\kern0pt}p{\kern0pt}l{\kern0pt}u{\kern0pt}g{\kern0pt}i{\kern0pt}n{\kern0pt}s{\kern0pt}/{\kern0pt}v{\kern0pt}d{\kern0pt}m{\kern0pt}2{\kern0pt}i{\kern0pt}s{\kern0pt}a{\kern0pt}/{\kern0pt}s{\kern0pt}r{\kern0pt}c{\kern0pt}/{\kern0pt}m{\kern0pt}a{\kern0pt}i{\kern0pt}n{\kern0pt}/{\kern0pt}r{\kern0pt}e{\kern0pt}s{\kern0pt}o{\kern0pt}u{\kern0pt}r{\kern0pt}c{\kern0pt}e{\kern0pt}s{\kern0pt}/{\kern0pt}R{\kern0pt}e{\kern0pt}c{\kern0pt}u{\kern0pt}r{\kern0pt}s{\kern0pt}i{\kern0pt}v{\kern0pt}e{\kern0pt}V{\kern0pt}D{\kern0pt}M{\kern0pt}.{\kern0pt}*{\kern0pt}}\footnote{in~\url{https://github.com/leouk/VDM_Toolkit}}.

In the next section, we present background on VDM and Isabelle recursion and
measure relations. Section~\ref{sec:VDMTypes} briefly discuss VDM basic
types translation and their consequence for recursion. Next,
Section~\ref{sec:Recursion} describes how both VDM and Isabelle recursive
definitions work and how they differ. Our translation strategy is then
presented in Section~\ref{sec:Strategy} for basic types, sets, maps, and
complex recursive patterns. Finally, we conclude in
Section~\ref{sec:Conclusion}.%
\end{isamarkuptext}\isamarkuptrue%
\isadelimdocument
\endisadelimdocument
\isatagdocument
\isamarkupsection{Background\label{sec:Background}%
}
\isamarkuptrue%
\endisatagdocument
{\isafolddocument}%
\isadelimdocument
\endisadelimdocument
\begin{isamarkuptext}%
Our VDM to Isabelle/HOL translator caters for a wide range of the VDM-SL
AST. It copes with all kinds of expressions, a variety of patterns, almost all
types, imports and exports, functions and specifications, traces, and some of
state and operations \cite{NimFull,AdvancedVSCodePaper}. Although not
all VDM patterns are allowed, the translator copes with most, and
where it does not, there is a corresponding equivalent.

One particular area we want to extend translation is over recursively defined
functions. VDM requires the user to define a measure function to justify why
recursion terminates. It then generates proof obligations to ensure
totality and termination.

Finally, our translation strategy follows the size-change termination (SCT)
proof strategy described in \cite{SCT_POPL,SCNP_POPL}. In particular, its
SCP (polynomial) and SCNP (non-polynomial) subclasses of recursive definitions
within the SCT, which permits efficient termination checking. Effectively, if
every infinite computation would give rise to an infinitely decreasing value
sequence (according to the size-change principle), then no infinite
computation is possible. Termination problems in this class have a global
ranking function of a certain form, which can be found using SAT solving,
hence increasing automation.%
\end{isamarkuptext}\isamarkuptrue%
\isadelimdocument
\endisadelimdocument
\isatagdocument
\isamarkupsection{VDM basic types in Isabelle\label{sec:VDMTypes}%
}
\isamarkuptrue%
\endisatagdocument
{\isafolddocument}%
\isadelimdocument
\endisadelimdocument
\begin{isamarkuptext}%
Isabelle represents natural numbers (\isa{{\isasymnat}}) as a (data) type with
two constructors (\isa{{\isadigit{0}}} and \isa{Suc\ n}), where all numbers are
projections over such constructions (\emph{e.g.}~\isa{{\isadigit{2}}\ {\isacharequal}{\kern0pt}\ Suc\ {\isacharparenleft}{\kern0pt}Suc\ {\isadigit{0}}{\isacharparenright}{\kern0pt}}). Isabelle integers (\isa{{\isasymint}}) are defined as a quotient type
involving two natural numbers. Isabelle quotient types are injections into a
constructively defined type. As with integers, other Isabelle numeric types
(\emph{e.g.,}~rationals \isa{{\isasymrat}}, reals \isa{{\isasymreal}}, \emph{etc}.) are defined in
terms of some involved natural number construction. Type conversions (or
coercions) are then defined to allow users to jump between type spaces.
Nevertheless, Isabelle has no implicit type widening rule for \isa{{\isasymnat}};
instead, it takes conventions like \isa{{\isacharparenleft}{\kern0pt}{\isadigit{0}}{\isacharcolon}{\kern0pt}{\isacharcolon}{\kern0pt}{\isasymnat}{\isacharparenright}{\kern0pt}\ {\isacharminus}{\kern0pt}\ {\isacharparenleft}{\kern0pt}x{\isacharcolon}{\kern0pt}{\isacharcolon}{\kern0pt}{\isasymnat}{\isacharparenright}{\kern0pt}\ {\isacharequal}{\kern0pt}\ {\isacharparenleft}{\kern0pt}{\isadigit{0}}{\isacharcolon}{\kern0pt}{\isacharcolon}{\kern0pt}{\isasymnat}{\isacharparenright}{\kern0pt}}. For expressions involving a mixutre of \isa{{\isasymint}} and \isa{{\isasymnat}} typed
terms, explicit user-defined type coercions might be needed
(\emph{e.g.,}~\isa{int\ {\isacharparenleft}{\kern0pt}{\isadigit{2}}{\isacharcolon}{\kern0pt}{\isacharcolon}{\kern0pt}{\isasymnat}{\isacharparenright}{\kern0pt}\ {\isacharminus}{\kern0pt}\ {\isacharparenleft}{\kern0pt}{\isadigit{3}}{\isacharcolon}{\kern0pt}{\isacharcolon}{\kern0pt}{\isasymint}{\isacharparenright}{\kern0pt}\ {\isacharequal}{\kern0pt}\ {\isacharminus}{\kern0pt}\ {\isacharparenleft}{\kern0pt}{\isadigit{1}}{\isacharcolon}{\kern0pt}{\isacharcolon}{\kern0pt}{\isasymint}{\isacharparenright}{\kern0pt}}).
 
VDM expressions with basic-typed (\textbf{nat}, \textbf{int}) variables have specific
type widening rules. For example, even if both variables are \textbf{nat}, the
result might be \textbf{int}. (\emph{e.g.,}~in VDM, \isatt{0{\kern0pt}\ {\char`\-}{\kern0pt}\ x{\kern0pt}:{\kern0pt}n{\kern0pt}a{\kern0pt}t{\kern0pt}\ ={\kern0pt}\ {\char`\-}{\kern0pt}x{\kern0pt}:{\kern0pt}i{\kern0pt}n{\kern0pt}t{\kern0pt}}). Therefore,
our translation strategy considers VDM \textbf{nat} as the Isabelle type
\isa{VDMNat}, which is just a type synonym for \isa{{\isasymint}}. This simplifies
the translation process to Isabelle, such that no type coercions are
necessary to encode all VDM type widenning rules. On the other hand, this
design decision means encoding of recursive functions over \textbf{nat} to be more
complicated than expected, given VDM's \textbf{nat} is represented as Isabelle's
\isa{{\isasymint}}.

  Despite this design decision over basic types and their consequences,
  recursion over VDM \textbf{int}, sets or maps will still be involved. That is
  because these types are not constructively defined in Isabelle.%
\end{isamarkuptext}\isamarkuptrue%
\isadelimdocument
\endisadelimdocument
\isatagdocument
\isamarkupsection{Recursion in VDM and in Isabelle\label{sec:Recursion}%
}
\isamarkuptrue%
\endisatagdocument
{\isafolddocument}%
\isadelimdocument
\endisadelimdocument
\begin{isamarkuptext}%
An important aspect of every recursive definition is an argument that
justifies its termination. Otherwise, the recursion might go into an infinite
loop. In VDM, this is defined using a recursive measure:~it has the same input
type signature as the recursive definition, and returns a \textbf{nat}, which
\textbf{must} monotonically decrease at each recursive call, eventually reaching
zero. This is how termination of recursive definitions are justified in VDM. A
simple example of a VDM recursive definition is one for calculating the
factorial of a given natural number
\begin{vdmsl}[frame=none,basicstyle=\ttfamily\scriptsize]
   fact: nat -> nat 
   fact(n) == if n = 0 then 1 else n * fact(n - 1)   measure n;   
\end{vdmsl} 
%
\noindent The \isatt{f{\kern0pt}a{\kern0pt}c{\kern0pt}t{\kern0pt}} recursive measure uses the \isatt{n{\kern0pt}} input itself as its
result. This works because the only recursive call is made with a decreasing
value of \isatt{n{\kern0pt}}, until it reaches the base case \isatt{0{\kern0pt}} and terminates. VDMJ generates three
proof obligations for the definition above. They are trivial to discharge in
Isabelle given the measure definition expanded is just 
\isa{{\isasymforall}n{\isachardot}{\kern0pt}\ n\ {\isasymnoteq}\ {\isadigit{0}}\ {\isasymlongrightarrow}\ {\isadigit{0}}\ {\isasymle}\ n\ {\isacharminus}{\kern0pt}\ {\isadigit{1}}} and 
\isa{{\isasymforall}n{\isachardot}{\kern0pt}\ n\ {\isasymnoteq}\ {\isadigit{0}}\ {\isasymlongrightarrow}\ n\ {\isacharminus}{\kern0pt}\ {\isadigit{1}}\ {\isacharless}{\kern0pt}\ n}.
\begin{vdmsl}[frame=none,basicstyle=\ttfamily\scriptsize]
  Proof Obligation 1: (Unproved) fact; measure_fact: total function obligation 
  (forall n:nat & is_(measure_fact(n), nat))
  
  Proof Obligation 2: (Unproved) fact: subtype obligation 
  (forall n:nat & (not (n = 0) => (n - 1) >= 0))
  
  Proof Obligation 3: (Unproved) fact: recursive function obligation 
  (forall n:nat & (not (n=0) => measure_fact(n) > measure_fact((n-1))))
\end{vdmsl} 
%
Moreover, even though measures over recursive type structures are impossible
to define in VDM, they are easily described in Isabelle. For example, it is
not possible to write a measure in VDM over a recursive record defining a
linked list, such as \isatt{R{\kern0pt}\ :{\kern0pt}:{\kern0pt}\ v{\kern0pt}:{\kern0pt}\ n{\kern0pt}a{\kern0pt}t{\kern0pt}\ n{\kern0pt}:{\kern0pt}\ R{\kern0pt}}. This is automatically generated for
our representation of VDM records in Isabelle as a datatype. Other complex
recursive patterns are hard/impossible to represent in VDM (see
Section~\ref{subsec:Complex}).
  
In Isabelle, recursive definitions can be provided through primitive recursion
over inputs that are constructively defined, or more general function
definitions that produce proof obligations. The former insists on definition
for each type constructor and only provides simplification rules; whereas the
latter allows for more sophisticated input patterns and provides
simplification, elimination and induction rules, as well as partial function
considerations. For the purposes of this paper, we only consider function
definitions. Readers can find more about these differences in~\cite{IsaFunctionPackage}.

Isabelle's recursive functions require a proof obligation that its parameters
represent a constructive and compatible pattern, and that recursive calls
terminate. Constructive patterns covers all constructors in the data type
being used in the recursion inputs (\emph{i.e.,}~one equation for each of the
constructors of \isa{{\isasymnat}}, hence one involving \isa{{\isadigit{0}}} and another
involving \isa{Suc\ n}). Compatible patterns cover the multiple ways
patterns can be constructed will boil down to the pattern completeness cases
(\emph{e.g.,}~\isa{n\ {\isacharplus}{\kern0pt}\ {\isadigit{2}}} being simply multiple calls over defined
constructors like \isa{Suc\ {\isacharparenleft}{\kern0pt}Suc\ n{\isacharparenright}{\kern0pt}}). This is important to ensure that
recursion is well structured (\emph{i.e.,}~recursive calls will not get stuck
because some constructs are not available). For example, if you miss the
\isa{{\isadigit{0}}} case, eventually the \isa{Suc\ n} case will reach zero and
fail, as no patterns for zero exist. The proof obligation for termination
establishes that the recursion is well-founded. This has to be proved whenever
properties of the defined function are meant to be total.

Isabelle's function definitions can be given with either \isa{fun} or
\isa{function} syntax. The former attempts to automatically prove the
pattern constructive and compatible proofs and finds a measure for the
termination proof obligation. The latter requires the user to do these proofs
manually by providing a measure relation. It is better suited for cases where
\isa{fun} declarations fail, which usually involve complex or ill-defined
recursion.

The termination relation must be well-founded, which means have a well-ordered
induction principle over a partially ordered relation defined as\footnote{Details on 
well-ordered induction are in Isabelle's \isatt{W{\kern0pt}e{\kern0pt}l{\kern0pt}l{\kern0pt}f{\kern0pt}o{\kern0pt}u{\kern0pt}n{\kern0pt}d{\kern0pt}e{\kern0pt}d{\kern0pt}.{\kern0pt}t{\kern0pt}h{\kern0pt}y{\kern0pt}} theory.} 
\begin{isabelle}%
wf\ {\isacharparenleft}{\kern0pt}r{\isacharcolon}{\kern0pt}{\isacharcolon}{\kern0pt}{\isacharparenleft}{\kern0pt}{\isacharprime}{\kern0pt}a\ {\isasymtimes}\ {\isacharprime}{\kern0pt}a{\isacharparenright}{\kern0pt}\ set{\isacharparenright}{\kern0pt}\ {\isacharequal}{\kern0pt}\isanewline
{\isacharparenleft}{\kern0pt}{\isasymforall}P{\isacharcolon}{\kern0pt}{\isacharcolon}{\kern0pt}{\isacharprime}{\kern0pt}a\ {\isasymRightarrow}\ {\isasymbool}{\isachardot}{\kern0pt}\ {\isacharparenleft}{\kern0pt}{\isasymforall}x{\isacharcolon}{\kern0pt}{\isacharcolon}{\kern0pt}{\isacharprime}{\kern0pt}a{\isachardot}{\kern0pt}\ {\isacharparenleft}{\kern0pt}{\isasymforall}y{\isacharcolon}{\kern0pt}{\isacharcolon}{\kern0pt}{\isacharprime}{\kern0pt}a{\isachardot}{\kern0pt}\ {\isacharparenleft}{\kern0pt}y{\isacharcomma}{\kern0pt}\ x{\isacharparenright}{\kern0pt}\ {\isasymin}\ r\ {\isasymlongrightarrow}\ P\ y{\isacharparenright}{\kern0pt}\ {\isasymlongrightarrow}\ P\ x{\isacharparenright}{\kern0pt}\ {\isasymlongrightarrow}\ {\isacharparenleft}{\kern0pt}{\isasymforall}x{\isacharcolon}{\kern0pt}{\isacharcolon}{\kern0pt}{\isacharprime}{\kern0pt}a{\isachardot}{\kern0pt}\ P\ x{\isacharparenright}{\kern0pt}{\isacharparenright}{\kern0pt}%
\end{isabelle} A definition that
Isabelle discovers all three proofs is%
\end{isamarkuptext}\isamarkuptrue%
\isacommand{fun}\isamarkupfalse%
\ fact{\isacharprime}{\kern0pt}\ {\isacharcolon}{\kern0pt}{\isacharcolon}{\kern0pt}\ {\isacartoucheopen}{\isasymnat}\ {\isasymRightarrow}\ {\isasymnat}{\isacartoucheclose}\ \isakeyword{where}\ {\isacartoucheopen}fact{\isacharprime}{\kern0pt}\ n\ {\isacharequal}{\kern0pt}\ {\isacharparenleft}{\kern0pt}if\ n\ {\isacharequal}{\kern0pt}\ {\isadigit{0}}\ then\ {\isadigit{1}}\ else\ n\ {\isacharasterisk}{\kern0pt}\ {\isacharparenleft}{\kern0pt}fact{\isacharprime}{\kern0pt}\ {\isacharparenleft}{\kern0pt}n\ {\isacharminus}{\kern0pt}\ {\isadigit{1}}{\isacharparenright}{\kern0pt}{\isacharparenright}{\kern0pt}{\isacharparenright}{\kern0pt}{\isacartoucheclose}%
\begin{isamarkuptext}%
This definition is quite similar in VDM. Nevertheless, VDM basic types
widening rules necessitate we translate them to \isa{VDMNat}. The same
version of \isa{fact} defined for \isa{{\isasymint}} will fail with the error that
``\emph{Could not find lexicographic termination order}''. That is, Isabelle
manages to discharge the pattern proofs, but not the termination one. This is
because the user must provide a projection relation from the \isa{{\isasymint}}
quotient type back into the constructive type \isa{{\isasymnat}}.

Even if we could avoid this translation technicality, the same problem would
occur for recursion over non-constructive types, such as sets or maps. They
require recursion over finite sets, which are defined inductively. The only
easy recursive translations are those involving lists, given lists in Isabelle
are defined constructively and VDM sequences map directly.

Therefore, defining recursive functions over non-constructive types entail
compatibility and completeness proofs. They also lead to partial function
definitions, given Isabelle cannot tell whether termination is immediately
obvious. In VDM, however, recursive functions on sets (as well as map domains)
are common, hence the need for extending our translation strategy.%
\end{isamarkuptext}\isamarkuptrue%
\isadelimdocument
\endisadelimdocument
\isatagdocument
\isamarkupsection{VDM recursion translation strategy\label{sec:Strategy}%
}
\isamarkuptrue%
\endisatagdocument
{\isafolddocument}%
\isadelimdocument
\endisadelimdocument
\begin{isamarkuptext}%
We want to identify a translation strategy that will cater for 
issues described above not only for basic types, but also for sets, sequences,
maps, as well as mutual recursion.

The VDM-SL AST tags all recursive functions, even those without an explicit
measure. All such functions will be translated using Isabelle's \isa{fun}
syntax, which will attempt to discover proofs for compatibility and
termination. For our setup of the \isatt{f{\kern0pt}a{\kern0pt}c{\kern0pt}t{\kern0pt}} example, where Isabelle discovers the
termination proof.%
\end{isamarkuptext}\isamarkuptrue%
\begin{isamarkuptext}%
If that had failed, the user could define a VDM \isatt{@{\kern0pt}I{\kern0pt}s{\kern0pt}a{\kern0pt}M{\kern0pt}e{\kern0pt}a{\kern0pt}s{\kern0pt}u{\kern0pt}r{\kern0pt}e{\kern0pt}}
annotation. VDM annotations are comments that will be processed according to
specific implementations \cite{VDMJAnnotations,AdvancedVSCodePaper}. If
the user does not provide an \isatt{@{\kern0pt}I{\kern0pt}s{\kern0pt}a{\kern0pt}M{\kern0pt}e{\kern0pt}a{\kern0pt}s{\kern0pt}u{\kern0pt}r{\kern0pt}e{\kern0pt}} annotation and \isa{fun}
fails, then it is up to the user to figure out the necessary proof setup.

The \isatt{@{\kern0pt}I{\kern0pt}s{\kern0pt}a{\kern0pt}M{\kern0pt}e{\kern0pt}a{\kern0pt}s{\kern0pt}u{\kern0pt}r{\kern0pt}e{\kern0pt}} annotation defines a well-founded measure relation that
will participate in the setup for Isabelle termination proof. For example, for
the \isatt{f{\kern0pt}a{\kern0pt}c{\kern0pt}t{\kern0pt}} example, the user would have to write an annotation before the VDM
measure.
\begin{vdmsl}[frame=none,basicstyle=\ttfamily\scriptsize]
  --@IsaMeasure( { (n -1, n) | n : nat & n <> 0 } )
\end{vdmsl}
\noindent This measure relation corresponds to the relationship between the
recursive call (\isatt{f{\kern0pt}a{\kern0pt}c{\kern0pt}t{\kern0pt}({\kern0pt}n{\kern0pt}{\char`\-}{\kern0pt}1{\kern0pt}){\kern0pt}}) and its defining equation (\isatt{f{\kern0pt}a{\kern0pt}c{\kern0pt}t{\kern0pt}({\kern0pt}n{\kern0pt}){\kern0pt}}), where
the filtering condition determines which values of \isatt{n{\kern0pt}} the relation should 
refer to. More interesting measure relations are defined in
Section~\ref{subsec:Complex}.

During translation, the \isatt{v{\kern0pt}d{\kern0pt}m{\kern0pt}2{\kern0pt}i{\kern0pt}s{\kern0pt}a{\kern0pt}} plugin will typecheck the \isatt{@{\kern0pt}I{\kern0pt}s{\kern0pt}a{\kern0pt}M{\kern0pt}e{\kern0pt}a{\kern0pt}s{\kern0pt}u{\kern0pt}r{\kern0pt}e{\kern0pt}} annotation
(\emph{i.e.,}~it is a type correct relation over the function signature). Next, it
will translate the annotation and some automation lemmas as a series of
Isabelle definitions to be used during the proof of termination of translated
VDM recursive functions. If no annotation is provided, following similar
principles from Isabelle, then the plugin will try to automatically infer what
the measure relation should be based on the structure of the recursive
function definition. If this fails, then the user is informed. Still, even
if measure-relation synthesis succeeds, the user still has to appropriately
use it during Isabelle's termination proof.

In what follows, we will detail the translation strategy for each relevant VDM
type. For details of the overall translation strategy, see examples in the
distribution\footnote{\url{https://github.com/leouk/VDM_Toolkit}} and~\cite{NimFull}.
The translation strategy imposes various implicit VDM checks as explicit predicates. 
For example, VDM sets are always finite, and type invariants over set, sequence
and map elements must hold for every element.%
\end{isamarkuptext}\isamarkuptrue%
\isadelimdocument
\endisadelimdocument
\isatagdocument
\isamarkupsubsection{Recursion over VDM basic types (\textbf{nat}, \textbf{int})\label{subsec:VDMNat}%
}
\isamarkuptrue%
\endisatagdocument
{\isafolddocument}%
\isadelimdocument
\endisadelimdocument
\begin{isamarkuptext}%
Following the general translation strategy~\cite{NimFull}, we first
encode the implicit precondition, which insists that
the given parameter \isa{n} is a \isa{VDMNat}, alongside a list of
defining constants that are useful for proof strategy synthesis.%
\end{isamarkuptext}\isamarkuptrue%
\isacommand{definition}\isamarkupfalse%
\ pre{\isacharunderscore}{\kern0pt}fact\ {\isacharcolon}{\kern0pt}{\isacharcolon}{\kern0pt}\ {\isacartoucheopen}VDMNat\ {\isasymRightarrow}\ {\isasymbool}{\isacartoucheclose}\ \isakeyword{where}\ {\isacartoucheopen}pre{\isacharunderscore}{\kern0pt}fact\ n\ {\isasymequiv}\ inv{\isacharunderscore}{\kern0pt}VDMNat\ n{\isacartoucheclose}\isanewline
\isacommand{lemmas}\isamarkupfalse%
\ pre{\isacharunderscore}{\kern0pt}fact{\isacharunderscore}{\kern0pt}defs\ {\isacharequal}{\kern0pt}\ pre{\isacharunderscore}{\kern0pt}fact{\isacharunderscore}{\kern0pt}def%
\begin{isamarkuptext}%
Next, we define factorial recursively. When the precondition fails, we
return \isa{undefined}, which is a term that cannot be reasoned with in
Isabelle The \isa{domintros} tag tells Isabelle to generate domain
predicates, in case this function is not total. Domain predicates are
important to our strategy because every VDM function will be undefined, when
applied outside its precondition. It also generates domain-predicate sensitive
proof rules listed below.%
\end{isamarkuptext}\isamarkuptrue%
\isacommand{function}\isamarkupfalse%
\ {\isacharparenleft}{\kern0pt}domintros{\isacharparenright}{\kern0pt}\ fact\ {\isacharcolon}{\kern0pt}{\isacharcolon}{\kern0pt}\ {\isacartoucheopen}VDMNat\ {\isasymRightarrow}\ VDMNat{\isacartoucheclose}\ \isakeyword{where}\isanewline
{\isacartoucheopen}fact\ n\ {\isacharequal}{\kern0pt}\ {\isacharparenleft}{\kern0pt}if\ pre{\isacharunderscore}{\kern0pt}fact\ n\ then\ {\isacharparenleft}{\kern0pt}if\ n\ {\isacharequal}{\kern0pt}\ {\isadigit{0}}\ then\ {\isadigit{1}}\ else\ n\ {\isacharasterisk}{\kern0pt}\ {\isacharparenleft}{\kern0pt}fact\ {\isacharparenleft}{\kern0pt}n\ {\isacharminus}{\kern0pt}\ {\isadigit{1}}{\isacharparenright}{\kern0pt}{\isacharparenright}{\kern0pt}{\isacharparenright}{\kern0pt}\ else\ undefined{\isacharparenright}{\kern0pt}{\isacartoucheclose}%
\begin{isamarkuptext}%
The proof obligations for pattern compatibility and completeness are
discharged with the usual Isabelle proof strategy for simple recursive
patterns with the \isa{pat{\isacharunderscore}{\kern0pt}completeness} method. In the general case
discussed in Section~\ref{subsec:Complex}, the user might have goals to
discharge. Isabelle proves various theorems abou case analysis and
(partial) rules for elimination, induction and simplification.%
\end{isamarkuptext}\isamarkuptrue%
\isadeliminvisible
\ \ %
\endisadeliminvisible
\isataginvisible
\isacommand{by}\isamarkupfalse%
\ {\isacharparenleft}{\kern0pt}pat{\isacharunderscore}{\kern0pt}completeness{\isacharcomma}{\kern0pt}\ auto{\isacharparenright}{\kern0pt}\ %
\endisataginvisible
{\isafoldinvisible}%
\isadeliminvisible
\endisadeliminvisible
\begin{isamarkuptext}%
Partial rules require the domain predicate \isa{fact{\isacharunderscore}{\kern0pt}dom} as an assumption.
It represents a well-founded relation that ensures termination. That is, if
the user does not want (or knows how) to prove termination, such domain
predicates will follow every application of definitions,
hence documenting the requirement that such well-founded relation is still
missing.

If/when the termination proof is discharged, these p-rules can be simplified
into total rules that do not depend on a domain predicate, given a
well-founded relation has been provided. Domain predicates will complicate
user proofs, and also make proof strategy synthesis harder to automate.

Termination proof is discharged by establishing a well-founded relation
associated with the function recursive call(s) with respect to its
declaration. The \isatt{@{\kern0pt}I{\kern0pt}s{\kern0pt}a{\kern0pt}M{\kern0pt}e{\kern0pt}a{\kern0pt}s{\kern0pt}u{\kern0pt}r{\kern0pt}e{\kern0pt}} annotation is translated as an
Isabelle abbreviation. The filter that the function
precondition holds:~this is important to ensure the termination proof never
reaches the \isa{undefined} case. The other filter comes from the negated
test from the if-statement. More complex definitions will have more
involved filters. We use abbreviation instead of definition to avoid needing
to expand the defined term in proofs.%
\end{isamarkuptext}\isamarkuptrue%
\isacommand{abbreviation}\isamarkupfalse%
\ fact{\isacharunderscore}{\kern0pt}wf\ {\isacharcolon}{\kern0pt}{\isacharcolon}{\kern0pt}\ {\isacartoucheopen}{\isacharparenleft}{\kern0pt}VDMNat\ {\isasymtimes}\ VDMNat{\isacharparenright}{\kern0pt}\ set{\isacartoucheclose}\ \isakeyword{where}\isanewline
\ \ {\isacartoucheopen}fact{\isacharunderscore}{\kern0pt}wf\ {\isasymequiv}\ {\isacharbraceleft}{\kern0pt}\ {\isacharparenleft}{\kern0pt}n\ {\isacharminus}{\kern0pt}\ {\isadigit{1}}{\isacharcomma}{\kern0pt}\ n{\isacharparenright}{\kern0pt}\ {\isacharbar}{\kern0pt}\ n\ {\isachardot}{\kern0pt}\ pre{\isacharunderscore}{\kern0pt}fact\ n\ {\isasymand}\ n\ {\isasymnoteq}\ {\isadigit{0}}\ {\isacharbraceright}{\kern0pt}{\isacartoucheclose}%
\begin{isamarkuptext}%
Given \isatt{f{\kern0pt}a{\kern0pt}c{\kern0pt}t{\kern0pt}} definition is simple (non-mutual, single call-site, easy
measure relation choice) recursion, its setup is easy to
establish well-foundedness. For recursions of this nature, we can piggyback on
Isabelle machinery to prove well foundedness by using terms
\isa{gen{\isacharunderscore}{\kern0pt}VDMNat{\isacharunderscore}{\kern0pt}term} and \isa{int{\isacharunderscore}{\kern0pt}ge{\isacharunderscore}{\kern0pt}less{\isacharunderscore}{\kern0pt}than}.

The first term is defined in terms of the second, which is a subset of our
well-founded relation \isa{fact{\isacharunderscore}{\kern0pt}wf}. Isabelle has proofs about the 
well foundedness of \isa{int{\isacharunderscore}{\kern0pt}ge{\isacharunderscore}{\kern0pt}less{\isacharunderscore}{\kern0pt}than}. Thus, making the proof our
term being well-founded trivial for \isacommand{sledgehammer}. 
As part of the translation strategy, we define
(and attempt to discover the proof of) the following lemma. This follows
the strategy described in~\cite{KrausSCNP}.%
\end{isamarkuptext}\isamarkuptrue%
\isacommand{lemma}\isamarkupfalse%
\ l{\isacharunderscore}{\kern0pt}fact{\isacharunderscore}{\kern0pt}term{\isacharunderscore}{\kern0pt}wf{\isacharcolon}{\kern0pt}\ {\isacartoucheopen}wf\ {\isacharparenleft}{\kern0pt}gen{\isacharunderscore}{\kern0pt}VDMNat{\isacharunderscore}{\kern0pt}term\ fact{\isacharunderscore}{\kern0pt}wf{\isacharparenright}{\kern0pt}{\isacartoucheclose}\isanewline
\isadeliminvisible
\ \ %
\endisadeliminvisible
\isataginvisible
\isacommand{by}\isamarkupfalse%
\ {\isacharparenleft}{\kern0pt}simp\ add{\isacharcolon}{\kern0pt}\ wf{\isacharunderscore}{\kern0pt}int{\isacharunderscore}{\kern0pt}ge{\isacharunderscore}{\kern0pt}less{\isacharunderscore}{\kern0pt}than\ wf{\isacharunderscore}{\kern0pt}Int{\isadigit{1}}{\isacharparenright}{\kern0pt}\ %
\endisataginvisible
{\isafoldinvisible}%
\isadeliminvisible
\endisadeliminvisible
\begin{isamarkuptext}%
Finally, we prove termination using the previously proved lemma using
the \isa{relation}. This simplifies the goal into well-foundedness of
termination relation and that the precondition implies it, both of which are
easily proved with simplification in this case.%
\end{isamarkuptext}\isamarkuptrue%
\isadeliminvisible
\endisadeliminvisible
\isataginvisible
\isacommand{termination}\isamarkupfalse%
\isanewline
\ \ \isacommand{apply}\isamarkupfalse%
\ {\isacharparenleft}{\kern0pt}relation\ {\isacartoucheopen}{\isacharparenleft}{\kern0pt}gen{\isacharunderscore}{\kern0pt}VDMNat{\isacharunderscore}{\kern0pt}term\ fact{\isacharunderscore}{\kern0pt}wf{\isacharparenright}{\kern0pt}{\isacartoucheclose}{\isacharparenright}{\kern0pt}\ %
\begin{isamarkuptext}%
This transforms the abstract domain predicate into two new subgoals as
\begin{isabelle}%
\ {\isadigit{1}}{\isachardot}{\kern0pt}\ wf\ {\isacharparenleft}{\kern0pt}gen{\isacharunderscore}{\kern0pt}VDMNat{\isacharunderscore}{\kern0pt}term\ fact{\isacharunderscore}{\kern0pt}wf{\isacharparenright}{\kern0pt}\isanewline
\ {\isadigit{2}}{\isachardot}{\kern0pt}\ {\isasymAnd}n{\isachardot}{\kern0pt}\ {\isasymlbrakk}pre{\isacharunderscore}{\kern0pt}fact\ n{\isacharsemicolon}{\kern0pt}\ n\ {\isasymnoteq}\ {\isadigit{0}}{\isasymrbrakk}\ {\isasymLongrightarrow}\ {\isacharparenleft}{\kern0pt}n\ {\isacharminus}{\kern0pt}\ {\isadigit{1}}{\isacharcomma}{\kern0pt}\ n{\isacharparenright}{\kern0pt}\ {\isasymin}\ gen{\isacharunderscore}{\kern0pt}VDMNat{\isacharunderscore}{\kern0pt}term\ fact{\isacharunderscore}{\kern0pt}wf%
\end{isabelle}%
\end{isamarkuptext}\isamarkuptrue%
\ \ \isacommand{using}\isamarkupfalse%
\ l{\isacharunderscore}{\kern0pt}fact{\isacharunderscore}{\kern0pt}term{\isacharunderscore}{\kern0pt}wf\ \isacommand{apply}\isamarkupfalse%
\ presburger\ \isanewline
\ \ \isacommand{by}\isamarkupfalse%
\ {\isacharparenleft}{\kern0pt}simp\ add{\isacharcolon}{\kern0pt}\ pre{\isacharunderscore}{\kern0pt}fact{\isacharunderscore}{\kern0pt}defs\ int{\isacharunderscore}{\kern0pt}ge{\isacharunderscore}{\kern0pt}less{\isacharunderscore}{\kern0pt}than{\isacharunderscore}{\kern0pt}def{\isacharparenright}{\kern0pt}\ %
\endisataginvisible
{\isafoldinvisible}%
\isadeliminvisible
\endisadeliminvisible
\begin{isamarkuptext}%
For this example, subgoals are proved with \isacommand{sledgehammer}. In general, the user will be have to either find the proof, or
deal with domain predicates involving the recursive call. After the
termination proof is discharged, Isabelle provides versions of rules for
elimination, induction and simplification that are total and do not depend on
the domain predicates. 

To make sure our choice does not lead to the empty relation, we ensure that
the termination relation is in fact the same as the well founded
predicate by proving the next goal. This is something users might want to do,
but is not part of the translation strategy. In case the measure relation is
empty, the recursive call simplification rules will not be useful anyhow.%
\end{isamarkuptext}\isamarkuptrue%
\isacommand{lemma}\isamarkupfalse%
\ l{\isacharunderscore}{\kern0pt}fact{\isacharunderscore}{\kern0pt}term{\isacharunderscore}{\kern0pt}valid{\isacharcolon}{\kern0pt}\ {\isacartoucheopen}{\isacharparenleft}{\kern0pt}gen{\isacharunderscore}{\kern0pt}VDMNat{\isacharunderscore}{\kern0pt}term\ fact{\isacharunderscore}{\kern0pt}wf{\isacharparenright}{\kern0pt}\ {\isacharequal}{\kern0pt}\ fact{\isacharunderscore}{\kern0pt}wf{\isacartoucheclose}\isanewline
\isadeliminvisible
\ \ %
\endisadeliminvisible
\isataginvisible
\isacommand{apply}\isamarkupfalse%
\ {\isacharparenleft}{\kern0pt}simp\ {\isacharparenright}{\kern0pt}\ \isanewline
\ \ \isacommand{apply}\isamarkupfalse%
\ {\isacharparenleft}{\kern0pt}intro\ equalityI\ subsetI{\isacharparenright}{\kern0pt}\isanewline
\ \ \ \isacommand{apply}\isamarkupfalse%
\ {\isacharparenleft}{\kern0pt}simp{\isacharunderscore}{\kern0pt}all\ add{\isacharcolon}{\kern0pt}\ pre{\isacharunderscore}{\kern0pt}fact{\isacharunderscore}{\kern0pt}defs\ int{\isacharunderscore}{\kern0pt}ge{\isacharunderscore}{\kern0pt}less{\isacharunderscore}{\kern0pt}than{\isacharunderscore}{\kern0pt}def\ case{\isacharunderscore}{\kern0pt}prod{\isacharunderscore}{\kern0pt}beta{\isacharparenright}{\kern0pt}\isanewline
\ \ \isacommand{by}\isamarkupfalse%
\ auto%
\endisataginvisible
{\isafoldinvisible}%
\isadeliminvisible
\endisadeliminvisible
\isadelimdocument
\endisadelimdocument
\isatagdocument
\isamarkupsubsection{Recursion over VDM sets\label{subsec:VDMSet}%
}
\isamarkuptrue%
\endisatagdocument
{\isafolddocument}%
\isadelimdocument
\endisadelimdocument
\begin{isamarkuptext}%
Next, we extend the translation strategy for VDM sets.
For this, we use a function that sums the elements of a set.
\begin{vdmsl}[frame=none,basicstyle=\ttfamily\scriptsize]
  sumset: set of nat -> nat 
  sumset(s) == if s = {} then 0 else let e in set s in sumset(s - {e}) + e
  --@IsaMeasure({(x - { let e in set x in e }, x) | x : set of nat & x <> {}}) 
  --@Witness(sumset({ 1 }))
  measure card s;
\end{vdmsl}
\noindent Most common VDM recursion over sets, consume the
set by picking an arbitrary set element and then recurse without
the element picked, until the set is empty. The VDM measure states that the
recursion is based on the cardinality of the input parameter. VDM measures are
not suitable for Isabelle proofs, given Isabelle requires a relation;~hence,
VDM measures are mostly ignored. They might still be useful during
proofs as potential witnesses to existentially-quantified goals.

The implicit VDM checks are defined as the precondition, which
ensures that the given set only contains natural numbers and is finite, as
defined by \isa{inv{\isacharunderscore}{\kern0pt}VDMSet{\isacharprime}{\kern0pt}}.%
\end{isamarkuptext}\isamarkuptrue%
\isacommand{definition}\isamarkupfalse%
\ pre{\isacharunderscore}{\kern0pt}sumset\ {\isacharcolon}{\kern0pt}{\isacharcolon}{\kern0pt}\ {\isacartoucheopen}VDMNat\ VDMSet\ {\isasymRightarrow}\ {\isasymbool}{\isacartoucheclose}\ \isakeyword{where}\isanewline
\ \ {\isacartoucheopen}pre{\isacharunderscore}{\kern0pt}sumset\ s\ {\isasymequiv}\ inv{\isacharunderscore}{\kern0pt}VDMSet{\isacharprime}{\kern0pt}\ inv{\isacharunderscore}{\kern0pt}VDMNat\ s{\isacartoucheclose}\isanewline
\isacommand{lemmas}\isamarkupfalse%
\ pre{\isacharunderscore}{\kern0pt}sumset{\isacharunderscore}{\kern0pt}defs\ {\isacharequal}{\kern0pt}\ pre{\isacharunderscore}{\kern0pt}sumset{\isacharunderscore}{\kern0pt}def\ inv{\isacharunderscore}{\kern0pt}VDMSet{\isacharprime}{\kern0pt}{\isacharunderscore}{\kern0pt}defs%
\begin{isamarkuptext}%
We define the VDM recursive function in Isabelle next. It checks whether
the given set satisfy the function precondition, returning \isa{undefined}
if not. Each case is encoded pretty much 1-1 from VDM after that. The
translation strategy for VDM \isatt{l{\kern0pt}e{\kern0pt}t{\kern0pt}{\char`\-}{\kern0pt}i{\kern0pt}n{\kern0pt}{\char`\-}{\kern0pt}s{\kern0pt}e{\kern0pt}t{\kern0pt}} patterns uses Isabelle's Hilbert's
choice operator (\isa{SOME\ x{\isachardot}{\kern0pt}\ x\ {\isasymin}\ s}). Note this naturally extends to VDM's
\isatt{l{\kern0pt}e{\kern0pt}t{\kern0pt}{\char`\-}{\kern0pt}b{\kern0pt}e{\kern0pt}{\char`\-}{\kern0pt}s{\kern0pt}t{\kern0pt}} patterns as well.%
\end{isamarkuptext}\isamarkuptrue%
\isacommand{function}\isamarkupfalse%
\ {\isacharparenleft}{\kern0pt}domintros{\isacharparenright}{\kern0pt}\ sumset\ {\isacharcolon}{\kern0pt}{\isacharcolon}{\kern0pt}\ {\isacartoucheopen}VDMNat\ VDMSet\ {\isasymRightarrow}\ VDMNat{\isacartoucheclose}\ \isakeyword{where}\ \isanewline
\ \ {\isacartoucheopen}sumset\ s\ {\isacharequal}{\kern0pt}\ {\isacharparenleft}{\kern0pt}if\ pre{\isacharunderscore}{\kern0pt}sumset\ s\ then\ \isanewline
\ \ \ \ \ \ \ \ \ \ \ \ \ \ \ \ \ \ \ \ \ \ \ {\isacharparenleft}{\kern0pt}if\ s\ {\isacharequal}{\kern0pt}\ {\isacharbraceleft}{\kern0pt}{\isacharbraceright}{\kern0pt}\ then\ {\isadigit{0}}\ else\ let\ e\ {\isacharequal}{\kern0pt}\ {\isacharparenleft}{\kern0pt}{\isasymsome}\ x\ {\isachardot}{\kern0pt}\ x\ {\isasymin}\ s{\isacharparenright}{\kern0pt}\ in\ sumset\ {\isacharparenleft}{\kern0pt}s\ {\isacharminus}{\kern0pt}\ {\isacharbraceleft}{\kern0pt}e{\isacharbraceright}{\kern0pt}{\isacharparenright}{\kern0pt}\ {\isacharplus}{\kern0pt}\ e{\isacharparenright}{\kern0pt}\ \isanewline
\ \ \ \ \ \ \ \ \ \ \ \ \ \ \ \ \ \ \ \ else\ undefined{\isacharparenright}{\kern0pt}{\isacartoucheclose}\isanewline
\isadeliminvisible
\ \ %
\endisadeliminvisible
\isataginvisible
\isacommand{by}\isamarkupfalse%
\ {\isacharparenleft}{\kern0pt}pat{\isacharunderscore}{\kern0pt}completeness{\isacharcomma}{\kern0pt}\ auto{\isacharparenright}{\kern0pt}\ %
\endisataginvisible
{\isafoldinvisible}%
\isadeliminvisible
\endisadeliminvisible
\begin{isamarkuptext}%
The pattern completeness and compatibility goals are trivial.
The measure relation defined with  \isatt{@{\kern0pt}I{\kern0pt}s{\kern0pt}a{\kern0pt}M{\kern0pt}e{\kern0pt}a{\kern0pt}s{\kern0pt}u{\kern0pt}r{\kern0pt}e{\kern0pt}} is next.
It is defined as the smaller set after picking \isa{e} and the set used 
at the entry call, leading to the pairs \isa{{\isacharparenleft}{\kern0pt}s\ {\isacharminus}{\kern0pt}\ {\isacharbraceleft}{\kern0pt}SOME\ e{\isachardot}{\kern0pt}\ e\ {\isasymin}\ s{\isacharbraceright}{\kern0pt}{\isacharcomma}{\kern0pt}\ s{\isacharparenright}{\kern0pt}}. 
Finally, we ensure all the relation elements satisfy the function precondition
and that the if-test is negated.%
\end{isamarkuptext}\isamarkuptrue%
\isacommand{abbreviation}\isamarkupfalse%
\ sumset{\isacharunderscore}{\kern0pt}wf{\isacharunderscore}{\kern0pt}rel\ {\isacharcolon}{\kern0pt}{\isacharcolon}{\kern0pt}\ {\isacartoucheopen}{\isacharparenleft}{\kern0pt}VDMNat\ VDMSet\ {\isasymtimes}\ VDMNat\ VDMSet{\isacharparenright}{\kern0pt}\ set{\isacartoucheclose}\ \isakeyword{where}\isanewline
\ \ {\isacartoucheopen}sumset{\isacharunderscore}{\kern0pt}wf{\isacharunderscore}{\kern0pt}rel\ {\isasymequiv}\ {\isacharbraceleft}{\kern0pt}\ {\isacharparenleft}{\kern0pt}s\ {\isacharminus}{\kern0pt}\ {\isacharbraceleft}{\kern0pt}{\isacharparenleft}{\kern0pt}{\isasymsome}\ e\ {\isachardot}{\kern0pt}\ e\ {\isasymin}\ s{\isacharparenright}{\kern0pt}{\isacharbraceright}{\kern0pt}{\isacharcomma}{\kern0pt}\ s{\isacharparenright}{\kern0pt}{\isacharbar}{\kern0pt}\ s\ {\isachardot}{\kern0pt}\ pre{\isacharunderscore}{\kern0pt}sumset\ s\ {\isasymand}\ s\ {\isasymnoteq}\ {\isacharbraceleft}{\kern0pt}{\isacharbraceright}{\kern0pt}{\isacharbraceright}{\kern0pt}{\isacartoucheclose}%
\begin{isamarkuptext}%
Given this is a simple recursion, again we can piggyback on Isabelle 
machinery by using the terms \isa{gen{\isacharunderscore}{\kern0pt}set{\isacharunderscore}{\kern0pt}term} and \isa{finite{\isacharunderscore}{\kern0pt}psubset}.
They establishes that a relation where the first element is strictly smaller
set than the second in the relation pair is well-founded. This makes
the proof of well-foundedness easy for \isacommand{sledgehammer}.%
\end{isamarkuptext}\isamarkuptrue%
\isacommand{lemma}\isamarkupfalse%
\ l{\isacharunderscore}{\kern0pt}sumset{\isacharunderscore}{\kern0pt}rel{\isacharunderscore}{\kern0pt}wf{\isacharcolon}{\kern0pt}\ {\isacartoucheopen}wf\ {\isacharparenleft}{\kern0pt}gen{\isacharunderscore}{\kern0pt}set{\isacharunderscore}{\kern0pt}term\ sumset{\isacharunderscore}{\kern0pt}wf{\isacharunderscore}{\kern0pt}rel{\isacharparenright}{\kern0pt}{\isacartoucheclose}\isanewline
\isadeliminvisible
\ \ %
\endisadeliminvisible
\isataginvisible
\isacommand{using}\isamarkupfalse%
\ l{\isacharunderscore}{\kern0pt}gen{\isacharunderscore}{\kern0pt}set{\isacharunderscore}{\kern0pt}term{\isacharunderscore}{\kern0pt}wf\ \isanewline
\ \ \isacommand{by}\isamarkupfalse%
\ blast\ %
\endisataginvisible
{\isafoldinvisible}%
\isadeliminvisible
\endisadeliminvisible
\begin{isamarkuptext}%
Next, we tackle the termination proof, with the same setup with \isa{relation} again.%
\end{isamarkuptext}\isamarkuptrue%
\isadeliminvisible
\endisadeliminvisible
\isataginvisible
\isacommand{termination}\isamarkupfalse%
\isanewline
\ \ \isacommand{apply}\isamarkupfalse%
\ {\isacharparenleft}{\kern0pt}relation\ {\isacartoucheopen}{\isacharparenleft}{\kern0pt}gen{\isacharunderscore}{\kern0pt}set{\isacharunderscore}{\kern0pt}term\ sumset{\isacharunderscore}{\kern0pt}wf{\isacharunderscore}{\kern0pt}rel{\isacharparenright}{\kern0pt}{\isacartoucheclose}{\isacharparenright}{\kern0pt}\isanewline
\ \ \isacommand{using}\isamarkupfalse%
\ l{\isacharunderscore}{\kern0pt}sumset{\isacharunderscore}{\kern0pt}rel{\isacharunderscore}{\kern0pt}wf\ \ \isanewline
\ \ \isacommand{apply}\isamarkupfalse%
\ force%
\begin{isamarkuptext}%
Unfortunately, using \isacommand{sledgehammer} fails to discharge the
  second subgoal %
\begin{isabelle}%
\ {\isadigit{1}}{\isachardot}{\kern0pt}\ {\isasymAnd}s\ x{\isachardot}{\kern0pt}\ {\isasymlbrakk}pre{\isacharunderscore}{\kern0pt}sumset\ s{\isacharsemicolon}{\kern0pt}\ s\ {\isasymnoteq}\ {\isasymemptyset}{\isacharsemicolon}{\kern0pt}\ x\ {\isacharequal}{\kern0pt}\ {\isacharparenleft}{\kern0pt}SOME\ x{\isachardot}{\kern0pt}\ x\ {\isasymin}\ s{\isacharparenright}{\kern0pt}{\isasymrbrakk}\isanewline
\isaindent{\ {\isadigit{1}}{\isachardot}{\kern0pt}\ {\isasymAnd}s\ x{\isachardot}{\kern0pt}\ }{\isasymLongrightarrow}\ {\isacharparenleft}{\kern0pt}s\ {\isacharminus}{\kern0pt}\ {\isacharbraceleft}{\kern0pt}x{\isacharbraceright}{\kern0pt}{\isacharcomma}{\kern0pt}\ s{\isacharparenright}{\kern0pt}\ {\isasymin}\ gen{\isacharunderscore}{\kern0pt}set{\isacharunderscore}{\kern0pt}term\ sumset{\isacharunderscore}{\kern0pt}wf{\isacharunderscore}{\kern0pt}rel%
\end{isabelle}%
\end{isamarkuptext}\isamarkuptrue%
\ \ \isacommand{oops}\isamarkupfalse%
\endisataginvisible
{\isafoldinvisible}%
\isadeliminvisible
\endisadeliminvisible
\begin{isamarkuptext}%
Fortunately, for most simple situations, this is easy to decompose in
general. The translation strategy takes the \isatt{@{\kern0pt}I{\kern0pt}s{\kern0pt}a{\kern0pt}M{\kern0pt}e{\kern0pt}a{\kern0pt}s{\kern0pt}u{\kern0pt}r{\kern0pt}e{\kern0pt}} expression and
decomposes its parts, such that the filtering predicates are assumptions, and
the element in the relation belongs to the well-founded measure chosen. This 
is defined in the next lemma, which require some manual intervention until 
Isabelle's \isacommand{sledgehammer} can finish the proof.%
\end{isamarkuptext}\isamarkuptrue%
\isacommand{lemma}\isamarkupfalse%
\ l{\isacharunderscore}{\kern0pt}pre{\isacharunderscore}{\kern0pt}sumset{\isacharunderscore}{\kern0pt}sumset{\isacharunderscore}{\kern0pt}wf{\isacharunderscore}{\kern0pt}rel{\isacharcolon}{\kern0pt}\ \isanewline
\ \ \ {\isacartoucheopen}pre{\isacharunderscore}{\kern0pt}sumset\ s\ {\isasymLongrightarrow}\ s\ {\isasymnoteq}\ {\isacharbraceleft}{\kern0pt}{\isacharbraceright}{\kern0pt}\ {\isasymLongrightarrow}\ {\isacharparenleft}{\kern0pt}s\ {\isacharminus}{\kern0pt}\ {\isacharbraceleft}{\kern0pt}{\isacharparenleft}{\kern0pt}{\isasymsome}\ x{\isachardot}{\kern0pt}\ x\ {\isasymin}\ s{\isacharparenright}{\kern0pt}{\isacharbraceright}{\kern0pt}{\isacharcomma}{\kern0pt}\ s{\isacharparenright}{\kern0pt}\ {\isasymin}\ {\isacharparenleft}{\kern0pt}gen{\isacharunderscore}{\kern0pt}set{\isacharunderscore}{\kern0pt}term\ sumset{\isacharunderscore}{\kern0pt}wf{\isacharunderscore}{\kern0pt}rel{\isacharparenright}{\kern0pt}{\isacartoucheclose}\isanewline
\isadeliminvisible
\ \ %
\endisadeliminvisible
\isataginvisible
\isacommand{unfolding}\isamarkupfalse%
\ gen{\isacharunderscore}{\kern0pt}set{\isacharunderscore}{\kern0pt}term{\isacharunderscore}{\kern0pt}def\ \isanewline
\ \ \isacommand{apply}\isamarkupfalse%
\ {\isacharparenleft}{\kern0pt}simp\ add{\isacharcolon}{\kern0pt}\ pre{\isacharunderscore}{\kern0pt}sumset{\isacharunderscore}{\kern0pt}defs{\isacharparenright}{\kern0pt}\isanewline
\ \ \isacommand{by}\isamarkupfalse%
\ {\isacharparenleft}{\kern0pt}metis\ Diff{\isacharunderscore}{\kern0pt}subset\ member{\isacharunderscore}{\kern0pt}remove\ psubsetI\ remove{\isacharunderscore}{\kern0pt}def\ some{\isacharunderscore}{\kern0pt}in{\isacharunderscore}{\kern0pt}eq{\isacharparenright}{\kern0pt}%
\endisataginvisible
{\isafoldinvisible}%
\isadeliminvisible
\endisadeliminvisible
\begin{isamarkuptext}%
The intuition behind this lemma is that elements in the measure
relation satisfy well-foundedness under the function precondition and the
filtering case (\isa{s\ {\isasymnoteq}\ {\isasymemptyset}}) where the recursive call is made. That is,
the precondition and filtering condition help establish the terminating
relation. For this particular proof, the only aspect needed from the
precondition is that the set is finite.
With this, we try the termination proof and \isacommand{sledgehammer} find proofs for all subgoals.%
\end{isamarkuptext}\isamarkuptrue%
\isadeliminvisible
\endisadeliminvisible
\isataginvisible
\isacommand{termination}\isamarkupfalse%
\isanewline
\ \ \isacommand{apply}\isamarkupfalse%
\ {\isacharparenleft}{\kern0pt}relation\ {\isacartoucheopen}{\isacharparenleft}{\kern0pt}gen{\isacharunderscore}{\kern0pt}set{\isacharunderscore}{\kern0pt}term\ sumset{\isacharunderscore}{\kern0pt}wf{\isacharunderscore}{\kern0pt}rel{\isacharparenright}{\kern0pt}{\isacartoucheclose}{\isacharparenright}{\kern0pt}\isanewline
\ \ \isacommand{using}\isamarkupfalse%
\ l{\isacharunderscore}{\kern0pt}sumset{\isacharunderscore}{\kern0pt}rel{\isacharunderscore}{\kern0pt}wf\ \isacommand{apply}\isamarkupfalse%
\ force\isanewline
\ \ \isacommand{using}\isamarkupfalse%
\ l{\isacharunderscore}{\kern0pt}pre{\isacharunderscore}{\kern0pt}sumset{\isacharunderscore}{\kern0pt}sumset{\isacharunderscore}{\kern0pt}wf{\isacharunderscore}{\kern0pt}rel\ \isacommand{by}\isamarkupfalse%
\ presburger%
\endisataginvisible
{\isafoldinvisible}%
\isadeliminvisible
\endisadeliminvisible
\isadelimproof
\endisadelimproof
\isatagproof
\endisatagproof
{\isafoldproof}%
\isadelimproof
\endisadelimproof
\begin{isamarkuptext}%
Note we omit such lemma over termination and precondition for the
\isa{VDMNat} case in Section~\ref{subsec:VDMNat}. The translation strategy
does define it following the same recipe, where \isacommand{sledgehammer} 
find the proof once more.%
\end{isamarkuptext}\isamarkuptrue%
\isacommand{lemma}\isamarkupfalse%
\ l{\isacharunderscore}{\kern0pt}pre{\isacharunderscore}{\kern0pt}fact{\isacharunderscore}{\kern0pt}wf{\isacharunderscore}{\kern0pt}rel{\isacharcolon}{\kern0pt}\ \isanewline
\ \ {\isacartoucheopen}{\isasymlbrakk}pre{\isacharunderscore}{\kern0pt}fact\ n{\isacharsemicolon}{\kern0pt}\ n\ {\isasymnoteq}\ {\isadigit{0}}{\isasymrbrakk}\ {\isasymLongrightarrow}\ {\isacharparenleft}{\kern0pt}n\ {\isacharminus}{\kern0pt}\ {\isadigit{1}}{\isacharcomma}{\kern0pt}\ n{\isacharparenright}{\kern0pt}\ {\isasymin}\ gen{\isacharunderscore}{\kern0pt}VDMNat{\isacharunderscore}{\kern0pt}term\ fact{\isacharunderscore}{\kern0pt}wf{\isacartoucheclose}\isanewline
\isadeliminvisible
\ \ %
\endisadeliminvisible
\isataginvisible
\isacommand{unfolding}\isamarkupfalse%
\ gen{\isacharunderscore}{\kern0pt}VDMNat{\isacharunderscore}{\kern0pt}term{\isacharunderscore}{\kern0pt}def\ gen{\isacharunderscore}{\kern0pt}VDMInt{\isacharunderscore}{\kern0pt}term{\isacharunderscore}{\kern0pt}def\ \isanewline
\ \ \isacommand{using}\isamarkupfalse%
\ l{\isacharunderscore}{\kern0pt}less{\isacharunderscore}{\kern0pt}than{\isacharunderscore}{\kern0pt}VDMNat{\isacharunderscore}{\kern0pt}subset{\isacharunderscore}{\kern0pt}int{\isacharunderscore}{\kern0pt}ge{\isacharunderscore}{\kern0pt}less{\isacharunderscore}{\kern0pt}than\ pre{\isacharunderscore}{\kern0pt}fact{\isacharunderscore}{\kern0pt}def\ \isanewline
\ \ \isacommand{by}\isamarkupfalse%
\ auto\ \isanewline
\endisataginvisible
{\isafoldinvisible}%
\isadeliminvisible
\endisadeliminvisible
\isadeliminvisible
\endisadeliminvisible
\isataginvisible
\endisataginvisible
{\isafoldinvisible}%
\isadeliminvisible
\endisadeliminvisible
\begin{isamarkuptext}%
Finally, even though this was not necessary for this proof, we encourage
users to always provide a witness for the top recursive call. This is done by
using the \isatt{@{\kern0pt}W{\kern0pt}i{\kern0pt}t{\kern0pt}n{\kern0pt}e{\kern0pt}s{\kern0pt}s{\kern0pt}} annotation~\cite{AdvancedVSCodePaper}:~it provides a 
concrete example for the function input parameters. This witness is useful 
for existentially quantified predicates of involved termination 
proofs (see Section~\ref{subsec:Complex}).%
\end{isamarkuptext}\isamarkuptrue%
\isadelimdocument
\endisadelimdocument
\isatagdocument
\isamarkupsubsection{Recursion over VDM maps\label{subsec:VDMMap}%
}
\isamarkuptrue%
\endisatagdocument
{\isafolddocument}%
\isadelimdocument
\endisadelimdocument
\begin{isamarkuptext}%
Recursive functions over maps are a special case of sets, given
map recursion usually iterates over the map's domain. For example, the
function that sums the elements of the map's range is defined as
\begin{vdmsl}[frame=none,basicstyle=\ttfamily\scriptsize]
  sum_elems: map nat to nat -> nat
  sum_elems(m) == 
    if m = {|->} then 0 else let d in set dom m in m(d) + sum_elems({d}<-:m)
  --@IsaMeasure({({d} <-: m, m) | m : map nat to nat, d: nat & 
                      m <> {} and d in set dom m})
  --@Witness( sum_elems({ 1 |-> 1 }) )
  measure card dom m;
\end{vdmsl}
\noindent As with sets, it iterates over the map by picking a domain element,
performing the necessary computation, and then recursing on the map filtered by
removing the chosen element, until the map is empty and the function
terminates. The measure relation follows the same
pattern:~recursive call site related with defining site, where both the
if-test and the \isatt{l{\kern0pt}e{\kern0pt}t{\kern0pt}{\char`\-}{\kern0pt}i{\kern0pt}n{\kern0pt}{\char`\-}{\kern0pt}s{\kern0pt}e{\kern0pt}t{\kern0pt}} choice is part of the filtering predicate.

Following the general translation strategy for maps~\cite{NimFull}, we define
the function precondition using \isa{inv{\isacharunderscore}{\kern0pt}Map}. It insists that both the map
domain and range are finite, and that all domain and range elements satisfy
their corresponding type invariant. Note that, if the recursion was defined
over sets other than the domain and range, then Isabelle require the proof
such set is finite. Given both domain and range sets are finite,
this should be easy, if needed.%
\end{isamarkuptext}\isamarkuptrue%
\isacommand{definition}\isamarkupfalse%
\ pre{\isacharunderscore}{\kern0pt}sum{\isacharunderscore}{\kern0pt}elems\ {\isacharcolon}{\kern0pt}{\isacharcolon}{\kern0pt}\ {\isacartoucheopen}{\isacharparenleft}{\kern0pt}VDMNat\ {\isasymrightharpoonup}\ VDMNat{\isacharparenright}{\kern0pt}\ {\isasymRightarrow}\ {\isasymbool}{\isacartoucheclose}\ \isakeyword{where}\isanewline
\ \ {\isacartoucheopen}pre{\isacharunderscore}{\kern0pt}sum{\isacharunderscore}{\kern0pt}elems\ m\ {\isasymequiv}\ inv{\isacharunderscore}{\kern0pt}VDMMap\ inv{\isacharunderscore}{\kern0pt}VDMNat\ inv{\isacharunderscore}{\kern0pt}VDMNat\ m{\isacartoucheclose}\isanewline
\isacommand{lemmas}\isamarkupfalse%
\ pre{\isacharunderscore}{\kern0pt}sum{\isacharunderscore}{\kern0pt}elems{\isacharunderscore}{\kern0pt}defs\ {\isacharequal}{\kern0pt}\ pre{\isacharunderscore}{\kern0pt}sum{\isacharunderscore}{\kern0pt}elems{\isacharunderscore}{\kern0pt}def\ inv{\isacharunderscore}{\kern0pt}VDMMap{\isacharunderscore}{\kern0pt}defs%
\begin{isamarkuptext}%
VDM maps in Isabelle (\isa{VDMNat\ {\isasymrightharpoonup}\ VDMNat}) are defined as a HOL
function which maps to an optional result. That is, if the element is in the
domain, then the map results in a non {\texttt{\textbf{nil}}} value; whereas, if the
element does not belong to the domain, then the map results in a {\texttt{\textbf{nil}}}
value. This makes all maps total, where values outside the domain
map to {\texttt{\textbf{nil}}}. The Isabelle translation and compatibility proof
follows patterns used before.%
\end{isamarkuptext}\isamarkuptrue%
\isacommand{function}\isamarkupfalse%
\ {\isacharparenleft}{\kern0pt}domintros{\isacharparenright}{\kern0pt}\ sum{\isacharunderscore}{\kern0pt}elems\ {\isacharcolon}{\kern0pt}{\isacharcolon}{\kern0pt}\ {\isacartoucheopen}{\isacharparenleft}{\kern0pt}VDMNat\ {\isasymrightharpoonup}\ VDMNat{\isacharparenright}{\kern0pt}\ {\isasymRightarrow}\ VDMNat{\isacartoucheclose}\ \isakeyword{where}\isanewline
\ \ {\isacartoucheopen}sum{\isacharunderscore}{\kern0pt}elems\ m\ {\isacharequal}{\kern0pt}\ \isanewline
\ \ \ \ {\isacharparenleft}{\kern0pt}if\ pre{\isacharunderscore}{\kern0pt}sum{\isacharunderscore}{\kern0pt}elems\ m\ then\isanewline
\ \ \ \ \ \ \ \ \ if\ m\ {\isacharequal}{\kern0pt}\ Map{\isachardot}{\kern0pt}empty\ then\ {\isadigit{0}}\ else\isanewline
\ \ \ \ \ \ \ \ \ \ \ \ \ let\ d\ {\isacharequal}{\kern0pt}\ {\isacharparenleft}{\kern0pt}{\isasymsome}\ e\ {\isachardot}{\kern0pt}\ e\ {\isasymin}\ dom\ m{\isacharparenright}{\kern0pt}\ in\ the{\isacharparenleft}{\kern0pt}m\ d{\isacharparenright}{\kern0pt}\ {\isacharplus}{\kern0pt}\ {\isacharparenleft}{\kern0pt}sum{\isacharunderscore}{\kern0pt}elems\ {\isacharparenleft}{\kern0pt}{\isacharbraceleft}{\kern0pt}d{\isacharbraceright}{\kern0pt}\ {\isacharminus}{\kern0pt}{\isasymtriangleleft}\ m{\isacharparenright}{\kern0pt}{\isacharparenright}{\kern0pt}\isanewline
\ \ \ \ \ else\ undefined{\isacharparenright}{\kern0pt}{\isacartoucheclose}\isanewline
\isadeliminvisible
\ \ %
\endisadeliminvisible
\isataginvisible
\isacommand{by}\isamarkupfalse%
\ {\isacharparenleft}{\kern0pt}pat{\isacharunderscore}{\kern0pt}completeness{\isacharcomma}{\kern0pt}\ auto{\isacharparenright}{\kern0pt}\ %
\endisataginvisible
{\isafoldinvisible}%
\isadeliminvisible
\endisadeliminvisible
\begin{isamarkuptext}%
Similarly, the well-founded relation is translated
next, where the precondition is included as part of the relation's
filter. The \isa{{\isacharbraceleft}{\kern0pt}d{\isacharbraceright}{\kern0pt}\ {\isacharminus}{\kern0pt}{\isasymtriangleleft}\ m} corresponds to the VDM domain anti-filtering operator
\isatt{{\char`\{}{\kern0pt}d{\kern0pt}{\char`\}}{\kern0pt}{\char`\<}{\kern0pt}{\char`\-}{\kern0pt}:{\kern0pt}m{\kern0pt}}.%
\end{isamarkuptext}\isamarkuptrue%
\isacommand{abbreviation}\isamarkupfalse%
\ \isanewline
\ \ sum{\isacharunderscore}{\kern0pt}elems{\isacharunderscore}{\kern0pt}wf\ {\isacharcolon}{\kern0pt}{\isacharcolon}{\kern0pt}\ {\isacartoucheopen}{\isacharparenleft}{\kern0pt}{\isacharparenleft}{\kern0pt}VDMNat\ {\isasymrightharpoonup}\ VDMNat{\isacharparenright}{\kern0pt}\ {\isasymtimes}\ {\isacharparenleft}{\kern0pt}VDMNat\ {\isasymrightharpoonup}\ VDMNat{\isacharparenright}{\kern0pt}{\isacharparenright}{\kern0pt}\ VDMSet{\isacartoucheclose}\ \isanewline
\ \ \isakeyword{where}\ {\isacartoucheopen}sum{\isacharunderscore}{\kern0pt}elems{\isacharunderscore}{\kern0pt}wf\ {\isasymequiv}\ \isanewline
\ \ {\isacharbraceleft}{\kern0pt}\ {\isacharparenleft}{\kern0pt}{\isacharparenleft}{\kern0pt}{\isacharbraceleft}{\kern0pt}d{\isacharbraceright}{\kern0pt}\ {\isacharminus}{\kern0pt}{\isasymtriangleleft}\ m{\isacharparenright}{\kern0pt}{\isacharcomma}{\kern0pt}\ m{\isacharparenright}{\kern0pt}\ {\isacharbar}{\kern0pt}\ m\ d\ {\isachardot}{\kern0pt}\ pre{\isacharunderscore}{\kern0pt}sum{\isacharunderscore}{\kern0pt}elems\ m\ {\isasymand}\ m\ {\isasymnoteq}\ Map{\isachardot}{\kern0pt}empty\ {\isasymand}\ d\ {\isasymin}\ dom\ m\ {\isacharbraceright}{\kern0pt}{\isacartoucheclose}%
\begin{isamarkuptext}%
For the well-founded lemma over the recursive measure relation, there is
no available Isabelle help, and projecting the domain element of the maps
within the relation is awkward. Thus, we have to prove the well-founded lemma 
and this will not be automatic in general. This is one difference in terms of
translation of VDM recursive functions over sets and maps.

Fortunately, the proof strategy for such situations is somewhat known:~it
follows a similar strategy to the proof of well foundedness of the
\isa{finite{\isacharunderscore}{\kern0pt}psubset}. It uses the provided VDM measure expression to extract 
the right projection of interest, then follows the proof for
 \isa{finite{\isacharunderscore}{\kern0pt}psubset}, where \isacommand{sledgehammer} can 
find the final steps.

The precondition subgoal and the termination proof follow the same
patterns as before. Their proof was discovered with \isacommand{sledgehammer}, yet this will not be the case in general.%
\end{isamarkuptext}\isamarkuptrue%
\isacommand{lemma}\isamarkupfalse%
\ l{\isacharunderscore}{\kern0pt}sum{\isacharunderscore}{\kern0pt}elems{\isacharunderscore}{\kern0pt}wf{\isacharcolon}{\kern0pt}\ {\isacartoucheopen}wf\ sum{\isacharunderscore}{\kern0pt}elems{\isacharunderscore}{\kern0pt}wf{\isacartoucheclose}\isanewline
\isadeliminvisible
\ \ %
\endisadeliminvisible
\isataginvisible
\isacommand{apply}\isamarkupfalse%
\ {\isacharparenleft}{\kern0pt}rule\ wf{\isacharunderscore}{\kern0pt}measure{\isacharbrackleft}{\kern0pt}of\ {\isacartoucheopen}{\isasymlambda}\ m\ {\isachardot}{\kern0pt}\ card\ {\isacharparenleft}{\kern0pt}dom\ m{\isacharparenright}{\kern0pt}{\isacartoucheclose}{\isacharcomma}{\kern0pt}\ THEN\ wf{\isacharunderscore}{\kern0pt}subset{\isacharbrackright}{\kern0pt}{\isacharparenright}{\kern0pt}\ \isanewline
\ \ \isacommand{apply}\isamarkupfalse%
\ {\isacharparenleft}{\kern0pt}simp\ add{\isacharcolon}{\kern0pt}\ measure{\isacharunderscore}{\kern0pt}def\ inv{\isacharunderscore}{\kern0pt}image{\isacharunderscore}{\kern0pt}def\ less{\isacharunderscore}{\kern0pt}than{\isacharunderscore}{\kern0pt}def\ less{\isacharunderscore}{\kern0pt}eq{\isacharparenright}{\kern0pt}\isanewline
\ \ \isacommand{apply}\isamarkupfalse%
\ {\isacharparenleft}{\kern0pt}rule\ subsetI{\isacharcomma}{\kern0pt}\ simp\ add{\isacharcolon}{\kern0pt}\ case{\isacharunderscore}{\kern0pt}prod{\isacharunderscore}{\kern0pt}beta{\isacharparenright}{\kern0pt}\isanewline
\ \ \isacommand{apply}\isamarkupfalse%
\ {\isacharparenleft}{\kern0pt}elim\ exE\ conjE{\isacharparenright}{\kern0pt}\isanewline
\ \ \isacommand{by}\isamarkupfalse%
\ {\isacharparenleft}{\kern0pt}simp\ add{\isacharcolon}{\kern0pt}\ l{\isacharunderscore}{\kern0pt}VDMMap{\isacharunderscore}{\kern0pt}filtering{\isacharunderscore}{\kern0pt}card\ pre{\isacharunderscore}{\kern0pt}sum{\isacharunderscore}{\kern0pt}elems{\isacharunderscore}{\kern0pt}defs{\isacharparenright}{\kern0pt}%
\endisataginvisible
{\isafoldinvisible}%
\isadeliminvisible
\isanewline
\endisadeliminvisible
\isanewline
\isacommand{lemma}\isamarkupfalse%
\ l{\isacharunderscore}{\kern0pt}pre{\isacharunderscore}{\kern0pt}sum{\isacharunderscore}{\kern0pt}elems{\isacharunderscore}{\kern0pt}sum{\isacharunderscore}{\kern0pt}elems{\isacharunderscore}{\kern0pt}wf{\isacharcolon}{\kern0pt}\ \isanewline
\ \ {\isacartoucheopen}{\isasymlbrakk}pre{\isacharunderscore}{\kern0pt}sum{\isacharunderscore}{\kern0pt}elems\ m{\isacharsemicolon}{\kern0pt}\ m\ {\isasymnoteq}\ Map{\isachardot}{\kern0pt}empty{\isasymrbrakk}\ {\isasymLongrightarrow}\ \isanewline
\ \ \ {\isacharparenleft}{\kern0pt}{\isacharbraceleft}{\kern0pt}{\isacharparenleft}{\kern0pt}{\isasymsome}\ e{\isachardot}{\kern0pt}\ e\ {\isasymin}\ dom\ m{\isacharparenright}{\kern0pt}{\isacharbraceright}{\kern0pt}\ {\isacharminus}{\kern0pt}{\isasymtriangleleft}\ m{\isacharcomma}{\kern0pt}\ m{\isacharparenright}{\kern0pt}\ {\isasymin}\ sum{\isacharunderscore}{\kern0pt}elems{\isacharunderscore}{\kern0pt}wf{\isacartoucheclose}\isanewline
\isadeliminvisible
\ \ %
\endisadeliminvisible
\isataginvisible
\isacommand{apply}\isamarkupfalse%
\ {\isacharparenleft}{\kern0pt}simp\ add{\isacharcolon}{\kern0pt}\ pre{\isacharunderscore}{\kern0pt}sum{\isacharunderscore}{\kern0pt}elems{\isacharunderscore}{\kern0pt}defs{\isacharparenright}{\kern0pt}\ \isanewline
\ \ \isacommand{by}\isamarkupfalse%
\ {\isacharparenleft}{\kern0pt}metis\ domIff\ empty{\isacharunderscore}{\kern0pt}iff\ some{\isacharunderscore}{\kern0pt}in{\isacharunderscore}{\kern0pt}eq{\isacharparenright}{\kern0pt}\ \isanewline
\isanewline
\isacommand{termination}\isamarkupfalse%
\isanewline
\ \ \isacommand{apply}\isamarkupfalse%
\ {\isacharparenleft}{\kern0pt}relation\ sum{\isacharunderscore}{\kern0pt}elems{\isacharunderscore}{\kern0pt}wf{\isacharparenright}{\kern0pt}\isanewline
\isanewline
\ \ \isacommand{apply}\isamarkupfalse%
\ {\isacharparenleft}{\kern0pt}simp\ add{\isacharcolon}{\kern0pt}\ l{\isacharunderscore}{\kern0pt}sum{\isacharunderscore}{\kern0pt}elems{\isacharunderscore}{\kern0pt}wf{\isacharparenright}{\kern0pt}\isanewline
\ \ \isacommand{using}\isamarkupfalse%
\ l{\isacharunderscore}{\kern0pt}pre{\isacharunderscore}{\kern0pt}sum{\isacharunderscore}{\kern0pt}elems{\isacharunderscore}{\kern0pt}sum{\isacharunderscore}{\kern0pt}elems{\isacharunderscore}{\kern0pt}wf\ \isacommand{by}\isamarkupfalse%
\ presburger%
\endisataginvisible
{\isafoldinvisible}%
\isadeliminvisible
\isanewline
\endisadeliminvisible
\isanewline
\isacommand{lemma}\isamarkupfalse%
\ l{\isacharunderscore}{\kern0pt}sum{\isacharunderscore}{\kern0pt}elems{\isacharunderscore}{\kern0pt}wf{\isacharunderscore}{\kern0pt}valid{\isacharcolon}{\kern0pt}\ {\isacartoucheopen}sum{\isacharunderscore}{\kern0pt}elems{\isacharunderscore}{\kern0pt}wf\ {\isasymnoteq}\ {\isacharbraceleft}{\kern0pt}{\isacharbraceright}{\kern0pt}{\isacartoucheclose}\isanewline
\isadeliminvisible
\ \ %
\endisadeliminvisible
\isataginvisible
\isacommand{apply}\isamarkupfalse%
\ safe\ \isanewline
\ \ \isacommand{apply}\isamarkupfalse%
\ {\isacharparenleft}{\kern0pt}erule\ equalityE{\isacharparenright}{\kern0pt}\isanewline
\ \ \isacommand{apply}\isamarkupfalse%
\ {\isacharparenleft}{\kern0pt}simp\ add{\isacharcolon}{\kern0pt}\ subset{\isacharunderscore}{\kern0pt}eq{\isacharparenright}{\kern0pt}\isanewline
\ \ \isacommand{apply}\isamarkupfalse%
\ {\isacharparenleft}{\kern0pt}erule{\isacharunderscore}{\kern0pt}tac\ x{\isacharequal}{\kern0pt}{\isacartoucheopen}{\isacharbrackleft}{\kern0pt}{\isadigit{1}}\ {\isasymmapsto}\ {\isadigit{1}}{\isacharbrackright}{\kern0pt}{\isacartoucheclose}\ \isakeyword{in}\ allE{\isacharparenright}{\kern0pt}\isanewline
\ \ \isacommand{by}\isamarkupfalse%
\ {\isacharparenleft}{\kern0pt}simp\ add{\isacharcolon}{\kern0pt}\ pre{\isacharunderscore}{\kern0pt}sum{\isacharunderscore}{\kern0pt}elems{\isacharunderscore}{\kern0pt}defs{\isacharparenright}{\kern0pt}\ %
\endisataginvisible
{\isafoldinvisible}%
\isadeliminvisible
\endisadeliminvisible
\begin{isamarkuptext}%
Finally, we also prove that the well founded termination relation
is not empty, as we did for sets and \textbf{nat}
recursion. Note that here the \isatt{@{\kern0pt}W{\kern0pt}i{\kern0pt}t{\kern0pt}n{\kern0pt}e{\kern0pt}s{\kern0pt}s{\kern0pt}} annotation is useful in discharging
the actual value to use as the witness demonstrating the relation is not
empty.%
\end{isamarkuptext}\isamarkuptrue%
\isadelimdocument
\endisadelimdocument
\isatagdocument
\isamarkupsubsection{VDM recursion involving complex measures\label{subsec:Complex}%
}
\isamarkuptrue%
\endisatagdocument
{\isafolddocument}%
\isadelimdocument
\endisadelimdocument
\begin{isamarkuptext}%
The class of recursive examples shown so far have covered a wide range of
situations, and have a good level of automation. Nevertheless, the same
strategy can also be applied for more complex recursive definitions. The cost
for the VDM user is the need of involved \isatt{@{\kern0pt}I{\kern0pt}s{\kern0pt}a{\kern0pt}M{\kern0pt}e{\kern0pt}a{\kern0pt}s{\kern0pt}u{\kern0pt}r{\kern0pt}e{\kern0pt}} definitions and
the highly likely need for extra user-defined lemmas. These lemmas can be
defined in VDM itself using the \isatt{@{\kern0pt}L{\kern0pt}e{\kern0pt}m{\kern0pt}m{\kern0pt}a{\kern0pt}} annotation.
To illustrate this, we define in VDM the (in)famous Ackermann
function\footnote{\url{https://en.wikipedia.org/wiki/Ackermann_function}}, which is a
staple example of complex recursion.
\begin{vdmsl}[frame=none,basicstyle=\ttfamily\scriptsize]
    ack: nat * nat -> nat 
    ack(m,n) == if m = 0 then n+1
           else if n = 0 then ack(m-1, 1)
           else               ack(m-1, ack(m, (n-1)))
    --@IsaMeasure( pair_less_VDMNat )
    --@Witness( ack(2, 1) )
    measure is not yet specified;
\end{vdmsl}
\noindent Note that the VDM measure is not defined, and that the
\isatt{@{\kern0pt}I{\kern0pt}s{\kern0pt}a{\kern0pt}M{\kern0pt}e{\kern0pt}a{\kern0pt}s{\kern0pt}u{\kern0pt}r{\kern0pt}e{\kern0pt}} uses a construct from Isabelle called \isa{pair{\isacharunderscore}{\kern0pt}less}. It
is part of Isabelle's machinery of concrete orders for SCNP problems \cite{KrausSCNP}. It considers recursions over multiple parameters, where some might
increase the number of calls (\emph{e.g.}~size-change). We are not aware of a
mechanism to define such measures in VDM.

We instantiate \isa{pair{\isacharunderscore}{\kern0pt}less} to \isa{VDMNat} as the
lexicographic product over the transitive closure of a totally ordered
relation between inputs\footnote{Details of this definition are in the
\isatt{V{\kern0pt}D{\kern0pt}M{\kern0pt}T{\kern0pt}o{\kern0pt}o{\kern0pt}l{\kern0pt}k{\kern0pt}i{\kern0pt}t{\kern0pt}.{\kern0pt}t{\kern0pt}h{\kern0pt}y{\kern0pt}} within the distribution.}.
If VDM measures were over relations, then the Ackermann measure could be defined
in VDM, assuming a standard definition of transitive closure\footnote{The
\isatt{R{\kern0pt}e{\kern0pt}l{\kern0pt}a{\kern0pt}t{\kern0pt}i{\kern0pt}o{\kern0pt}n{\kern0pt}s{\kern0pt}.{\kern0pt}v{\kern0pt}d{\kern0pt}m{\kern0pt}s{\kern0pt}l{\kern0pt}} provides such definition in the VDM toolkit distribution.}.
\begin{vdmsl}[frame=none,basicstyle=\ttfamily\scriptsize]
pair_less_VDMNat: () -> set of ((nat*nat) * (nat*nat))
pair_less_VDMNat() == lex_prod[nat, nat](less_than_VDMNat(), less_than_VDMNat());

less_than_VDMNat: () -> set of (nat*nat)
less_than_VDMNat() == trans_closure[nat]({ mk_(z', z) | z', z : nat & z' < z });

lex_prod[@A,@B]: set of (@A*@A) * set of (@B*@B) -> set of ((@A*@B) * (@A*@B))
lex_prod(ra,rb) == { mk_(mk_(a, b), mk_(a', b')) | a, a': @A, b, b': @B & 
                      mk_(a,a') in set ra or a = a' and mk_(b, b') in set rb };
\end{vdmsl}
That represents the lexicographic product of possibilities that are ordered in 
its parameters. Translation then outputs:%
\end{isamarkuptext}\isamarkuptrue%
\isacommand{definition}\isamarkupfalse%
\ pre{\isacharunderscore}{\kern0pt}ack\ {\isacharcolon}{\kern0pt}{\isacharcolon}{\kern0pt}\ {\isacartoucheopen}VDMNat\ {\isasymRightarrow}\ VDMNat\ {\isasymRightarrow}\ {\isasymbool}{\isacartoucheclose}\ \isakeyword{where}\isanewline
\ \ {\isacartoucheopen}pre{\isacharunderscore}{\kern0pt}ack\ m\ n\ {\isasymequiv}\ inv{\isacharunderscore}{\kern0pt}VDMNat\ m\ {\isasymand}\ inv{\isacharunderscore}{\kern0pt}VDMNat\ n{\isacartoucheclose}\isanewline
\isacommand{lemmas}\isamarkupfalse%
\ pre{\isacharunderscore}{\kern0pt}ack{\isacharunderscore}{\kern0pt}defs\ {\isacharequal}{\kern0pt}\ pre{\isacharunderscore}{\kern0pt}ack{\isacharunderscore}{\kern0pt}def\ \isanewline
\isanewline
\isacommand{function}\isamarkupfalse%
\ {\isacharparenleft}{\kern0pt}domintros{\isacharparenright}{\kern0pt}\ ack\ {\isacharcolon}{\kern0pt}{\isacharcolon}{\kern0pt}\ {\isacartoucheopen}VDMNat\ {\isasymRightarrow}\ VDMNat\ {\isasymRightarrow}\ VDMNat{\isacartoucheclose}\ \isakeyword{where}\isanewline
\ \ {\isacartoucheopen}ack\ m\ n\ {\isacharequal}{\kern0pt}\ {\isacharparenleft}{\kern0pt}if\ pre{\isacharunderscore}{\kern0pt}ack\ m\ n\ then\isanewline
\ \ \ \ \ \ \ \ \ \ \ \ \ \ \ \ \ \ \ \ \ \ \ if\ m\ {\isacharequal}{\kern0pt}\ {\isadigit{0}}\ then\ n{\isacharplus}{\kern0pt}{\isadigit{1}}\isanewline
\ \ \ \ \ \ \ \ \ \ \ \ \ \ \ \ \ \ else\ if\ n\ {\isacharequal}{\kern0pt}\ {\isadigit{0}}\ then\ ack\ {\isacharparenleft}{\kern0pt}m{\isacharminus}{\kern0pt}{\isadigit{1}}{\isacharparenright}{\kern0pt}\ {\isadigit{1}}\isanewline
\ \ \ \ \ \ \ \ \ \ \ \ \ \ \ \ \ \ else\ \ \ \ \ \ \ \ \ \ \ \ \ \ \ ack\ {\isacharparenleft}{\kern0pt}m{\isacharminus}{\kern0pt}{\isadigit{1}}{\isacharparenright}{\kern0pt}\ {\isacharparenleft}{\kern0pt}ack\ m\ {\isacharparenleft}{\kern0pt}n{\isacharminus}{\kern0pt}{\isadigit{1}}{\isacharparenright}{\kern0pt}{\isacharparenright}{\kern0pt}\isanewline
\ \ \ \ \ \ \ \ \ \ \ \ \ \ \ \ \ \ else\ \ \ \ \ \ \ \ \ \ \ \ \ \ \ undefined{\isacharparenright}{\kern0pt}{\isacartoucheclose}\isanewline
\isadeliminvisible
\ \ %
\endisadeliminvisible
\isataginvisible
\isacommand{by}\isamarkupfalse%
\ {\isacharparenleft}{\kern0pt}pat{\isacharunderscore}{\kern0pt}completeness{\isacharcomma}{\kern0pt}\ auto{\isacharparenright}{\kern0pt}\ %
\endisataginvisible
{\isafoldinvisible}%
\isadeliminvisible
\isanewline
\endisadeliminvisible
\isanewline
\isacommand{abbreviation}\isamarkupfalse%
\ ack{\isacharunderscore}{\kern0pt}wf\ {\isacharcolon}{\kern0pt}{\isacharcolon}{\kern0pt}\ {\isacartoucheopen}{\isacharparenleft}{\kern0pt}{\isacharparenleft}{\kern0pt}VDMNat\ {\isasymtimes}\ VDMNat{\isacharparenright}{\kern0pt}\ {\isasymtimes}\ {\isacharparenleft}{\kern0pt}VDMNat\ {\isasymtimes}\ VDMNat{\isacharparenright}{\kern0pt}{\isacharparenright}{\kern0pt}\ VDMSet{\isacartoucheclose}\ \isanewline
\ \ \isakeyword{where}\ {\isacartoucheopen}ack{\isacharunderscore}{\kern0pt}wf\ {\isasymequiv}\ pair{\isacharunderscore}{\kern0pt}less{\isacharunderscore}{\kern0pt}VDMNat{\isacartoucheclose}\isanewline
\isadeliminvisible
\isanewline
\endisadeliminvisible
\isataginvisible
\isacommand{termination}\isamarkupfalse%
\isanewline
\ \ \isacommand{apply}\isamarkupfalse%
\ {\isacharparenleft}{\kern0pt}relation\ ack{\isacharunderscore}{\kern0pt}wf{\isacharparenright}{\kern0pt}\isanewline
\ \ \isacommand{using}\isamarkupfalse%
\ wf{\isacharunderscore}{\kern0pt}pair{\isacharunderscore}{\kern0pt}less{\isacharunderscore}{\kern0pt}VDMNat\ \ \isacommand{apply}\isamarkupfalse%
\ blast\ \isanewline
\ \ \ \ \isacommand{apply}\isamarkupfalse%
\ {\isacharparenleft}{\kern0pt}simp\ add{\isacharcolon}{\kern0pt}\ l{\isacharunderscore}{\kern0pt}pair{\isacharunderscore}{\kern0pt}less{\isacharunderscore}{\kern0pt}VDMNat{\isacharunderscore}{\kern0pt}I{\isadigit{1}}\ pre{\isacharunderscore}{\kern0pt}ack{\isacharunderscore}{\kern0pt}def{\isacharparenright}{\kern0pt}\ \isanewline
\ \ \ \isacommand{apply}\isamarkupfalse%
\ {\isacharparenleft}{\kern0pt}simp\ add{\isacharcolon}{\kern0pt}\ \ pre{\isacharunderscore}{\kern0pt}ack{\isacharunderscore}{\kern0pt}def{\isacharparenright}{\kern0pt}\ \isanewline
\ \ \isacommand{by}\isamarkupfalse%
\ {\isacharparenleft}{\kern0pt}simp\ add{\isacharcolon}{\kern0pt}\ pair{\isacharunderscore}{\kern0pt}less{\isacharunderscore}{\kern0pt}VDMNat{\isacharunderscore}{\kern0pt}def\ pre{\isacharunderscore}{\kern0pt}ack{\isacharunderscore}{\kern0pt}def{\isacharparenright}{\kern0pt}\ %
\endisataginvisible
{\isafoldinvisible}%
\isadeliminvisible
\endisadeliminvisible
\begin{isamarkuptext}%
Compatibility and termination proofs are similar, despite the more 
complex measure, because of available Isabelle automation. We also show that 
this version of Ackermann with \isa{VDMNat} is equivalent to the usual Isabelle 
definition using \isa{{\isasymnat}}. We omit details here, but have proved that they are 
equivalent by induction.%
\end{isamarkuptext}\isamarkuptrue%
\isadelimproof
\endisadelimproof
\isatagproof
\endisatagproof
{\isafoldproof}%
\isadelimproof
\endisadelimproof
\isadelimproof
\endisadelimproof
\isatagproof
\endisatagproof
{\isafoldproof}%
\isadelimproof
\endisadelimproof
\isadelimproof
\endisadelimproof
\isatagproof
\endisatagproof
{\isafoldproof}%
\isadelimproof
\endisadelimproof
\isadelimproof
\endisadelimproof
\isatagproof
\endisatagproof
{\isafoldproof}%
\isadelimproof
\endisadelimproof
\isadelimproof
\endisadelimproof
\isatagproof
\endisatagproof
{\isafoldproof}%
\isadelimproof
\isanewline
\endisadelimproof
\isacommand{theorem}\isamarkupfalse%
\ ack{\isacharunderscore}{\kern0pt}correct{\isacharcolon}{\kern0pt}\ {\isacartoucheopen}ack{\isacharprime}{\kern0pt}\ m\ n\ {\isacharequal}{\kern0pt}\ ack\ m\ n{\isacartoucheclose}\isanewline
\isadelimproof
\ \ %
\endisadelimproof
\isatagproof
\isacommand{apply}\isamarkupfalse%
\ {\isacharparenleft}{\kern0pt}induction\ {\isacartoucheopen}m{\isacartoucheclose}\ {\isacartoucheopen}n{\isacartoucheclose}\ rule{\isacharcolon}{\kern0pt}\ ack{\isacharprime}{\kern0pt}{\isachardot}{\kern0pt}induct{\isacharparenright}{\kern0pt}\ \isacommand{by}\isamarkupfalse%
\ {\isacharparenleft}{\kern0pt}simp\ add{\isacharcolon}{\kern0pt}\ pre{\isacharunderscore}{\kern0pt}ack{\isacharunderscore}{\kern0pt}defs{\isacharparenright}{\kern0pt}{\isacharplus}{\kern0pt}%
\endisatagproof
{\isafoldproof}%
\isadelimproof
\endisadelimproof
\begin{isamarkuptext}%
In general, each complex recursive function will require such a setup.
Fortunately, Isabelle has a number of options available. Yet, in general, the
more complex the recursion, the more users will have to provide further
automation.%
\end{isamarkuptext}\isamarkuptrue%
\isadelimdocument
\endisadelimdocument
\isatagdocument
\isamarkupsubsection{Harder examples\label{subsec:Hard}%
}
\isamarkuptrue%
\endisatagdocument
{\isafolddocument}%
\isadelimdocument
\endisadelimdocument
\isadelimproof
\endisadelimproof
\isatagproof
\endisatagproof
{\isafoldproof}%
\isadelimproof
\endisadelimproof
\isadelimproof
\endisadelimproof
\isatagproof
\endisatagproof
{\isafoldproof}%
\isadelimproof
\endisadelimproof
\begin{isamarkuptext}%
We can handle all examples from~\cite{SCT_POPL}. We show here some that
require an elaborate setup. For example, the permutation function shows
permuting decreasing parameters with an involved measure. The precondition was
required for finishing the termination proof and shows an example why proofs
over \isa{{\isasymint}} can be harder. This illustrated how Isabelle (failed) proofs 
helped to improve the VDM specification. 
\begin{vdmsl}[frame=none,basicstyle=\ttfamily\scriptsize]
    perm: int * int * int -> int 
    perm(m,n,r) == if 0 < r then perm(m, r-1, n) 
              else if 0 < n then perm(r, n-1, m) else m
    --@IsaMeasure({mk_(mk_(m, r-1, n), mk_(m,n,r)) | ... & 0 < r} union 
    --            {mk_(mk_(r, n-1, m), mk_(m,n,r)) | ... & not 0 < r and 0 < n})
    pre ((0 < r or 0 < n) => m+n+r > 0)   measure maxs({m+n+r, 0});      

    tak: int * int * int -> int
    tak(x,y,z) == if x <= y then y  
                  else           tak(tak(x-1,y,z), tak(y-1,z,x), tak(z-1,x,y))
    measure is not yet specified;
\end{vdmsl}%
\end{isamarkuptext}\isamarkuptrue%
\isacommand{definition}\isamarkupfalse%
\ pre{\isacharunderscore}{\kern0pt}perm\ {\isacharcolon}{\kern0pt}{\isacharcolon}{\kern0pt}\ {\isacartoucheopen}VDMInt\ {\isasymRightarrow}\ VDMInt\ {\isasymRightarrow}\ VDMInt\ {\isasymRightarrow}\ {\isasymbool}{\isacartoucheclose}\ \isakeyword{where}\isanewline
\ \ {\isacartoucheopen}pre{\isacharunderscore}{\kern0pt}perm\ m\ n\ r\ {\isasymequiv}\ {\isacharparenleft}{\kern0pt}{\isacharparenleft}{\kern0pt}{\isadigit{0}}\ {\isacharless}{\kern0pt}\ r\ {\isasymor}\ {\isadigit{0}}\ {\isacharless}{\kern0pt}\ n{\isacharparenright}{\kern0pt}\ {\isasymlongrightarrow}\ m{\isacharplus}{\kern0pt}n{\isacharplus}{\kern0pt}r\ {\isachargreater}{\kern0pt}\ {\isadigit{0}}{\isacharparenright}{\kern0pt}{\isacartoucheclose}\isanewline
\isacommand{lemmas}\isamarkupfalse%
\ pre{\isacharunderscore}{\kern0pt}perm{\isacharunderscore}{\kern0pt}defs\ {\isacharequal}{\kern0pt}\ pre{\isacharunderscore}{\kern0pt}perm{\isacharunderscore}{\kern0pt}def\ inv{\isacharunderscore}{\kern0pt}VDMInt{\isacharunderscore}{\kern0pt}def\ inv{\isacharunderscore}{\kern0pt}True{\isacharunderscore}{\kern0pt}def\isanewline
\isadeliminvisible
\isanewline
\endisadeliminvisible
\isataginvisible
\isacommand{lemma}\isamarkupfalse%
\ l{\isacharunderscore}{\kern0pt}pre{\isacharunderscore}{\kern0pt}perm{\isacharunderscore}{\kern0pt}trivial{\isacharbrackleft}{\kern0pt}simp{\isacharbrackright}{\kern0pt}{\isacharcolon}{\kern0pt}\ {\isacartoucheopen}{\isacharparenleft}{\kern0pt}pre{\isacharunderscore}{\kern0pt}perm\ m\ n\ r{\isacharparenright}{\kern0pt}\ {\isacharequal}{\kern0pt}\ {\isacharparenleft}{\kern0pt}{\isacharparenleft}{\kern0pt}{\isadigit{0}}\ {\isacharless}{\kern0pt}\ r\ {\isasymor}\ {\isadigit{0}}\ {\isacharless}{\kern0pt}\ n{\isacharparenright}{\kern0pt}\ {\isasymlongrightarrow}\ m{\isacharplus}{\kern0pt}n{\isacharplus}{\kern0pt}r\ {\isachargreater}{\kern0pt}\ {\isadigit{0}}{\isacharparenright}{\kern0pt}{\isacartoucheclose}\ \isanewline
\ \ \isacommand{unfolding}\isamarkupfalse%
\ pre{\isacharunderscore}{\kern0pt}perm{\isacharunderscore}{\kern0pt}def\ inv{\isacharunderscore}{\kern0pt}VDMInt{\isacharunderscore}{\kern0pt}def\ \isanewline
\ \ \isacommand{by}\isamarkupfalse%
\ simp\ %
\endisataginvisible
{\isafoldinvisible}%
\isadeliminvisible
\isanewline
\endisadeliminvisible
\isanewline
\isacommand{function}\isamarkupfalse%
\ {\isacharparenleft}{\kern0pt}domintros{\isacharparenright}{\kern0pt}\ perm\ {\isacharcolon}{\kern0pt}{\isacharcolon}{\kern0pt}\ {\isacartoucheopen}VDMInt\ {\isasymRightarrow}\ VDMInt\ {\isasymRightarrow}\ VDMInt\ {\isasymRightarrow}\ VDMInt{\isacartoucheclose}\ \isakeyword{where}\isanewline
\ \ {\isacartoucheopen}perm\ m\ n\ r\ {\isacharequal}{\kern0pt}\ {\isacharparenleft}{\kern0pt}if\ pre{\isacharunderscore}{\kern0pt}perm\ m\ n\ r\ then\isanewline
\ \ \ \ \ \ \ \ \ \ \ \ \ \ \ \ \ \ \ \ \ \ \ \ \ \ \ \ \ \ \ \ if\ {\isadigit{0}}\ {\isacharless}{\kern0pt}\ r\ then\ perm\ m\ {\isacharparenleft}{\kern0pt}r{\isacharminus}{\kern0pt}{\isadigit{1}}{\isacharparenright}{\kern0pt}\ n\ \isanewline
\ \ \ \ \ \ \ \ \ \ \ \ \ \ \ \ \ \ \ \ \ \ \ \ \ \ \ else\ if\ {\isadigit{0}}\ {\isacharless}{\kern0pt}\ n\ then\ perm\ r\ {\isacharparenleft}{\kern0pt}n{\isacharminus}{\kern0pt}{\isadigit{1}}{\isacharparenright}{\kern0pt}\ m\ else\ m\isanewline
\ \ \ \ \ \ \ \ \ \ \ \ \ \ \ \ \ \ \ \ \ \ \ \ else\ undefined{\isacharparenright}{\kern0pt}{\isacartoucheclose}\isanewline
\isadeliminvisible
\ \ %
\endisadeliminvisible
\isataginvisible
\isacommand{by}\isamarkupfalse%
\ {\isacharparenleft}{\kern0pt}pat{\isacharunderscore}{\kern0pt}completeness{\isacharcomma}{\kern0pt}\ auto{\isacharparenright}{\kern0pt}\ %
\endisataginvisible
{\isafoldinvisible}%
\isadeliminvisible
\isanewline
\endisadeliminvisible
\isanewline
\isanewline
\isacommand{definition}\isamarkupfalse%
\ {\isacartoucheopen}perm{\isacharunderscore}{\kern0pt}wf{\isacharunderscore}{\kern0pt}rel\ {\isasymequiv}\ \isanewline
\ \ \ {\isacharbraceleft}{\kern0pt}\ {\isacharparenleft}{\kern0pt}{\isacharparenleft}{\kern0pt}m{\isacharcomma}{\kern0pt}\ r{\isacharminus}{\kern0pt}{\isadigit{1}}{\isacharcomma}{\kern0pt}\ n{\isacharparenright}{\kern0pt}{\isacharcomma}{\kern0pt}\ {\isacharparenleft}{\kern0pt}m{\isacharcomma}{\kern0pt}\ n{\isacharcomma}{\kern0pt}\ r{\isacharparenright}{\kern0pt}{\isacharparenright}{\kern0pt}\ {\isacharbar}{\kern0pt}\ m\ r\ n\ {\isachardot}{\kern0pt}\ pre{\isacharunderscore}{\kern0pt}perm\ m\ n\ r\ {\isasymand}\ {\isadigit{0}}\ {\isacharless}{\kern0pt}\ r\ {\isacharbraceright}{\kern0pt}\ {\isasymunion}\ \isanewline
\ \ \ {\isacharbraceleft}{\kern0pt}\ {\isacharparenleft}{\kern0pt}{\isacharparenleft}{\kern0pt}r{\isacharcomma}{\kern0pt}\ n{\isacharminus}{\kern0pt}{\isadigit{1}}{\isacharcomma}{\kern0pt}\ m{\isacharparenright}{\kern0pt}{\isacharcomma}{\kern0pt}\ {\isacharparenleft}{\kern0pt}m{\isacharcomma}{\kern0pt}\ n{\isacharcomma}{\kern0pt}\ r{\isacharparenright}{\kern0pt}{\isacharparenright}{\kern0pt}\ {\isacharbar}{\kern0pt}\ m\ r\ n\ {\isachardot}{\kern0pt}\ pre{\isacharunderscore}{\kern0pt}perm\ m\ n\ r\ {\isasymand}\ {\isasymnot}\ {\isadigit{0}}\ {\isacharless}{\kern0pt}\ r\ {\isasymand}\ {\isadigit{0}}\ {\isacharless}{\kern0pt}\ n\ {\isacharbraceright}{\kern0pt}{\isacartoucheclose}%
\begin{isamarkuptext}%
Its measure relation contains elements for each recursive call, filtered
for the corresponding if-then case. The proof of well-foundedness of such
measure relations involving multiple recursive calls require further proof
engineering, which is stated in the next lemma.%
\end{isamarkuptext}\isamarkuptrue%
\isacommand{lemma}\isamarkupfalse%
\ l{\isacharunderscore}{\kern0pt}perm{\isacharunderscore}{\kern0pt}wf{\isacharunderscore}{\kern0pt}rel{\isacharunderscore}{\kern0pt}VDM{\isacharunderscore}{\kern0pt}measure{\isacharcolon}{\kern0pt}\ \isanewline
\ \ {\isacartoucheopen}perm{\isacharunderscore}{\kern0pt}wf{\isacharunderscore}{\kern0pt}rel\ {\isasymsubseteq}\ measure\ {\isacharparenleft}{\kern0pt}{\isasymlambda}\ {\isacharparenleft}{\kern0pt}m{\isacharcomma}{\kern0pt}\ r{\isacharcomma}{\kern0pt}\ n{\isacharparenright}{\kern0pt}\ {\isachardot}{\kern0pt}\ nat\ {\isacharparenleft}{\kern0pt}max\ {\isadigit{0}}\ {\isacharparenleft}{\kern0pt}m{\isacharplus}{\kern0pt}r{\isacharplus}{\kern0pt}n{\isacharparenright}{\kern0pt}{\isacharparenright}{\kern0pt}{\isacharparenright}{\kern0pt}{\isacartoucheclose}\isanewline
\isadeliminvisible
\endisadeliminvisible
\isataginvisible
\isacommand{proof}\isamarkupfalse%
\ {\isacharminus}{\kern0pt}\ \isanewline
\ \ \ \isacommand{show}\isamarkupfalse%
\ {\isacharquery}{\kern0pt}thesis\isanewline
\ \ \isacommand{apply}\isamarkupfalse%
\ {\isacharparenleft}{\kern0pt}intro\ subsetI{\isacharcomma}{\kern0pt}\ case{\isacharunderscore}{\kern0pt}tac\ x{\isacharparenright}{\kern0pt}\isanewline
\ \ \ \ \isacommand{apply}\isamarkupfalse%
\ {\isacharparenleft}{\kern0pt}simp\ add{\isacharcolon}{\kern0pt}\ perm{\isacharunderscore}{\kern0pt}wf{\isacharunderscore}{\kern0pt}rel{\isacharunderscore}{\kern0pt}def\ case{\isacharunderscore}{\kern0pt}prod{\isacharunderscore}{\kern0pt}beta\ max{\isacharunderscore}{\kern0pt}def{\isacharparenright}{\kern0pt}\isanewline
\ \ \ \ \isacommand{apply}\isamarkupfalse%
\ {\isacharparenleft}{\kern0pt}elim\ disjE\ conjE{\isacharcomma}{\kern0pt}\ simp{\isacharparenright}{\kern0pt}\ \isanewline
\ \ \ \ \ \isacommand{apply}\isamarkupfalse%
\ {\isacharparenleft}{\kern0pt}intro\ impI\ conjI{\isacharcomma}{\kern0pt}\ simp{\isacharunderscore}{\kern0pt}all{\isacharparenright}{\kern0pt}\isanewline
\ \ \ \ \isacommand{done}\isamarkupfalse%
\isanewline
\isacommand{qed}\isamarkupfalse%
\endisataginvisible
{\isafoldinvisible}%
\isadeliminvisible
\isanewline
\endisadeliminvisible
\isanewline
\isacommand{lemma}\isamarkupfalse%
\ l{\isacharunderscore}{\kern0pt}perm{\isacharunderscore}{\kern0pt}wf{\isacharunderscore}{\kern0pt}rel{\isacharcolon}{\kern0pt}\ {\isacartoucheopen}wf\ perm{\isacharunderscore}{\kern0pt}wf{\isacharunderscore}{\kern0pt}rel{\isacartoucheclose}\isanewline
\isadeliminvisible
\endisadeliminvisible
\isataginvisible
\isacommand{proof}\isamarkupfalse%
\ {\isacharminus}{\kern0pt}\ \isanewline
\ \ \isacommand{from}\isamarkupfalse%
\ l{\isacharunderscore}{\kern0pt}perm{\isacharunderscore}{\kern0pt}wf{\isacharunderscore}{\kern0pt}rel{\isacharunderscore}{\kern0pt}VDM{\isacharunderscore}{\kern0pt}measure\ \isacommand{show}\isamarkupfalse%
\ {\isacharquery}{\kern0pt}thesis\isanewline
\ \ \isacommand{by}\isamarkupfalse%
\ {\isacharparenleft}{\kern0pt}rule\ wf{\isacharunderscore}{\kern0pt}subset\ {\isacharbrackleft}{\kern0pt}OF\ wf{\isacharunderscore}{\kern0pt}measure{\isacharbrackright}{\kern0pt}{\isacharparenright}{\kern0pt}\isanewline
\isacommand{qed}\isamarkupfalse%
\endisataginvisible
{\isafoldinvisible}%
\isadeliminvisible
\endisadeliminvisible
\begin{isamarkuptext}%
The Isabelle \isa{measure} definition projects the less-than ordered
inverse image\footnote{That is, \isa{measure\ {\isacharequal}{\kern0pt}\ inv{\isacharunderscore}{\kern0pt}image\ less{\isacharunderscore}{\kern0pt}than}. Inverse image is defined as \isa{inv{\isacharunderscore}{\kern0pt}image\ r\ f\ {\isacharequal}{\kern0pt}\ {\isacharbraceleft}{\kern0pt}{\isacharparenleft}{\kern0pt}x{\isacharcomma}{\kern0pt}\ y{\isacharparenright}{\kern0pt}\ {\isacharbar}{\kern0pt}\ {\isacharparenleft}{\kern0pt}f\ x{\isacharcomma}{\kern0pt}\ f\ y{\isacharparenright}{\kern0pt}\ {\isasymin}\ r{\isacharbraceright}{\kern0pt}}.} of a given function as the recursive measure relation. Here
the VDM-defined measure is given as such measure function projection. This
highlights to the VDM user the relationship (and differences) between VDM and
Isabelle recursive measures.

The setup works here if the \isa{pre{\isacharunderscore}{\kern0pt}perm} specifically curbs negative
sums of parameters. This was not immediately obvious. With the added
precondition the termination proof is discovered with \isacommand{sledgehammer}.%
\end{isamarkuptext}\isamarkuptrue%
\isadeliminvisible
\endisadeliminvisible
\isataginvisible
\isacommand{termination}\isamarkupfalse%
\isanewline
\ \ \isacommand{apply}\isamarkupfalse%
\ {\isacharparenleft}{\kern0pt}relation\ {\isacartoucheopen}perm{\isacharunderscore}{\kern0pt}wf{\isacharunderscore}{\kern0pt}rel{\isacartoucheclose}{\isacharparenright}{\kern0pt}\ \isanewline
\ \ \ \ \isacommand{apply}\isamarkupfalse%
\ {\isacharparenleft}{\kern0pt}simp\ add{\isacharcolon}{\kern0pt}\ l{\isacharunderscore}{\kern0pt}perm{\isacharunderscore}{\kern0pt}wf{\isacharunderscore}{\kern0pt}rel{\isacharparenright}{\kern0pt}\ \isanewline
\ \ \isacommand{by}\isamarkupfalse%
\ {\isacharparenleft}{\kern0pt}simp{\isacharunderscore}{\kern0pt}all\ add{\isacharcolon}{\kern0pt}\ perm{\isacharunderscore}{\kern0pt}wf{\isacharunderscore}{\kern0pt}rel{\isacharunderscore}{\kern0pt}def{\isacharparenright}{\kern0pt}\ \ %
\endisataginvisible
{\isafoldinvisible}%
\isadeliminvisible
\endisadeliminvisible
\begin{isamarkuptext}%
Finally, the Takeuchi's
function\footnote{\url{https://isabelle.in.tum.de/library/HOL/HOL-Examples/Functions.html}},
which contains both permutation and inner recursion is defined next, where the
important part is the SCNP setup using multi-sets \cite{KrausSCNP}. That is
needed because ordered lexicographic products are not strong enough to capture
this type of recursion. It has not implicit precondition. The translation
strategy works for these definitions, yet stands little chance of finding
proofs automatically.%
\end{isamarkuptext}\isamarkuptrue%
\isacommand{function}\isamarkupfalse%
\ {\isacharparenleft}{\kern0pt}domintros{\isacharparenright}{\kern0pt}\ tak\ {\isacharcolon}{\kern0pt}{\isacharcolon}{\kern0pt}\ {\isacartoucheopen}VDMInt\ {\isasymRightarrow}\ VDMInt\ {\isasymRightarrow}\ VDMInt\ {\isasymRightarrow}\ VDMInt{\isacartoucheclose}\ \isakeyword{where}\isanewline
\ \ {\isacartoucheopen}tak\ x\ y\ z\ {\isacharequal}{\kern0pt}\ {\isacharparenleft}{\kern0pt}if\ x\ {\isasymle}\ y\ then\ y\ else\ tak\ {\isacharparenleft}{\kern0pt}tak\ {\isacharparenleft}{\kern0pt}x{\isacharminus}{\kern0pt}{\isadigit{1}}{\isacharparenright}{\kern0pt}\ y\ z{\isacharparenright}{\kern0pt}\ {\isacharparenleft}{\kern0pt}tak\ {\isacharparenleft}{\kern0pt}y{\isacharminus}{\kern0pt}{\isadigit{1}}{\isacharparenright}{\kern0pt}\ z\ x{\isacharparenright}{\kern0pt}\ {\isacharparenleft}{\kern0pt}tak\ {\isacharparenleft}{\kern0pt}z{\isacharminus}{\kern0pt}{\isadigit{1}}{\isacharparenright}{\kern0pt}\ x\ y{\isacharparenright}{\kern0pt}{\isacharparenright}{\kern0pt}{\isacartoucheclose}\isanewline
\isadeliminvisible
\ \ %
\endisadeliminvisible
\isataginvisible
\isacommand{by}\isamarkupfalse%
\ auto\ %
\endisataginvisible
{\isafoldinvisible}%
\isadeliminvisible
\endisadeliminvisible
\begin{isamarkuptext}%
Next is an example of how one has to keep domain predicates as 
assumptions prior to termination proofs.%
\end{isamarkuptext}\isamarkuptrue%
\isacommand{lemma}\isamarkupfalse%
\ tak{\isacharunderscore}{\kern0pt}pcorrect{\isacharcolon}{\kern0pt}\isanewline
\ \ {\isacartoucheopen}tak{\isacharunderscore}{\kern0pt}dom\ {\isacharparenleft}{\kern0pt}x{\isacharcomma}{\kern0pt}\ y{\isacharcomma}{\kern0pt}\ z{\isacharparenright}{\kern0pt}\ {\isasymLongrightarrow}\ tak\ x\ y\ z\ {\isacharequal}{\kern0pt}\ {\isacharparenleft}{\kern0pt}if\ x\ {\isasymle}\ y\ then\ y\ else\ if\ y\ {\isasymle}\ z\ then\ z\ else\ x{\isacharparenright}{\kern0pt}{\isacartoucheclose}\isanewline
\isadeliminvisible
\ \ %
\endisadeliminvisible
\isataginvisible
\isacommand{thm}\isamarkupfalse%
\ tak{\isachardot}{\kern0pt}pinduct\ tak{\isachardot}{\kern0pt}psimps\ \isanewline
\ \ \isacommand{apply}\isamarkupfalse%
\ {\isacharparenleft}{\kern0pt}induction\ x\ y\ z\ rule{\isacharcolon}{\kern0pt}\ tak{\isachardot}{\kern0pt}pinduct{\isacharparenright}{\kern0pt}\ \isanewline
\ \ \isacommand{by}\isamarkupfalse%
\ {\isacharparenleft}{\kern0pt}simp\ add{\isacharcolon}{\kern0pt}\ tak{\isachardot}{\kern0pt}psimps{\isacharparenright}{\kern0pt}\ %
\endisataginvisible
{\isafoldinvisible}%
\isadeliminvisible
\endisadeliminvisible
\begin{isamarkuptext}%
Each case (including the non-recursive call) is represented in the SCNP
setup, and then their measure-lexicographic\footnote{The measure-lexicographic product
\isa{f\ {\isacharless}{\kern0pt}{\isacharasterisk}{\kern0pt}mlex{\isacharasterisk}{\kern0pt}{\isachargreater}{\kern0pt}\ R} is represented as the inverse image of the lexicographic product \isa{f\ {\isacharless}{\kern0pt}{\isacharasterisk}{\kern0pt}mlex{\isacharasterisk}{\kern0pt}{\isachargreater}{\kern0pt}\ R\ {\isacharequal}{\kern0pt}\ inv{\isacharunderscore}{\kern0pt}image\ {\isacharparenleft}{\kern0pt}less{\isacharunderscore}{\kern0pt}than\ {\isacharless}{\kern0pt}{\isacharasterisk}{\kern0pt}lex{\isacharasterisk}{\kern0pt}{\isachargreater}{\kern0pt}\ R{\isacharparenright}{\kern0pt}\ {\isacharparenleft}{\kern0pt}{\isasymlambda}x{\isachardot}{\kern0pt}\ {\isacharparenleft}{\kern0pt}f\ x{\isacharcomma}{\kern0pt}\ x{\isacharparenright}{\kern0pt}{\isacharparenright}{\kern0pt}}.} composition is used as the measure relation for the
termination proof.%
\end{isamarkuptext}\isamarkuptrue%
\isacommand{definition}\isamarkupfalse%
\ tak{\isacharunderscore}{\kern0pt}m{\isadigit{1}}\ \isakeyword{where}\ {\isacartoucheopen}tak{\isacharunderscore}{\kern0pt}m{\isadigit{1}}\ {\isacharequal}{\kern0pt}\ {\isacharparenleft}{\kern0pt}{\isasymlambda}{\isacharparenleft}{\kern0pt}x{\isacharcomma}{\kern0pt}y{\isacharcomma}{\kern0pt}z{\isacharparenright}{\kern0pt}{\isachardot}{\kern0pt}\ if\ x\ {\isasymle}\ y\ then\ {\isadigit{0}}\ else\ {\isadigit{1}}{\isacharparenright}{\kern0pt}{\isacartoucheclose}\isanewline
\isacommand{definition}\isamarkupfalse%
\ tak{\isacharunderscore}{\kern0pt}m{\isadigit{2}}\ \isakeyword{where}\ {\isacartoucheopen}tak{\isacharunderscore}{\kern0pt}m{\isadigit{2}}\ {\isacharequal}{\kern0pt}\ {\isacharparenleft}{\kern0pt}{\isasymlambda}{\isacharparenleft}{\kern0pt}x{\isacharcomma}{\kern0pt}y{\isacharcomma}{\kern0pt}z{\isacharparenright}{\kern0pt}{\isachardot}{\kern0pt}\ nat\ {\isacharparenleft}{\kern0pt}Max\ {\isacharbraceleft}{\kern0pt}x{\isacharcomma}{\kern0pt}\ y{\isacharcomma}{\kern0pt}\ z{\isacharbraceright}{\kern0pt}\ {\isacharminus}{\kern0pt}\ Min\ {\isacharbraceleft}{\kern0pt}x{\isacharcomma}{\kern0pt}\ y{\isacharcomma}{\kern0pt}\ z{\isacharbraceright}{\kern0pt}{\isacharparenright}{\kern0pt}{\isacharparenright}{\kern0pt}{\isacartoucheclose}\isanewline
\isacommand{definition}\isamarkupfalse%
\ tak{\isacharunderscore}{\kern0pt}m{\isadigit{3}}\ \isakeyword{where}\ {\isacartoucheopen}tak{\isacharunderscore}{\kern0pt}m{\isadigit{3}}\ {\isacharequal}{\kern0pt}\ {\isacharparenleft}{\kern0pt}{\isasymlambda}{\isacharparenleft}{\kern0pt}x{\isacharcomma}{\kern0pt}y{\isacharcomma}{\kern0pt}z{\isacharparenright}{\kern0pt}{\isachardot}{\kern0pt}\ nat\ {\isacharparenleft}{\kern0pt}x\ {\isacharminus}{\kern0pt}\ Min\ {\isacharbraceleft}{\kern0pt}x{\isacharcomma}{\kern0pt}\ y{\isacharcomma}{\kern0pt}\ z{\isacharbraceright}{\kern0pt}{\isacharparenright}{\kern0pt}{\isacharparenright}{\kern0pt}{\isacartoucheclose}\isanewline
\isadelimproof
\endisadelimproof
\isatagproof
\endisatagproof
{\isafoldproof}%
\isadelimproof
\endisadelimproof
\isadelimproof
\endisadelimproof
\isatagproof
\endisatagproof
{\isafoldproof}%
\isadelimproof
\endisadelimproof
\isadeliminvisible
\endisadeliminvisible
\isataginvisible
\endisataginvisible
{\isafoldinvisible}%
\isadeliminvisible
\endisadeliminvisible
\isadelimproof
\endisadelimproof
\isatagproof
\endisatagproof
{\isafoldproof}%
\isadelimproof
\endisadelimproof
\isadelimproof
\endisadelimproof
\isatagproof
\endisatagproof
{\isafoldproof}%
\isadelimproof
\endisadelimproof
\isadelimproof
\endisadelimproof
\isatagproof
\endisatagproof
{\isafoldproof}%
\isadelimproof
\endisadelimproof
\begin{isamarkuptext}%
The termination proof uses the measure relation \isa{tak{\isacharunderscore}{\kern0pt}m{\isadigit{1}}\ {\isacharless}{\kern0pt}{\isacharasterisk}{\kern0pt}mlex{\isacharasterisk}{\kern0pt}{\isachargreater}{\kern0pt}\ tak{\isacharunderscore}{\kern0pt}m{\isadigit{2}}\ {\isacharless}{\kern0pt}{\isacharasterisk}{\kern0pt}mlex{\isacharasterisk}{\kern0pt}{\isachargreater}{\kern0pt}\ tak{\isacharunderscore}{\kern0pt}m{\isadigit{3}}\ {\isacharless}{\kern0pt}{\isacharasterisk}{\kern0pt}mlex{\isacharasterisk}{\kern0pt}{\isachargreater}{\kern0pt}\ {\isasymemptyset}}. It requires user-defined lemmas for each
of the four cases.%
\end{isamarkuptext}\isamarkuptrue%
\isadeliminvisible
\endisadeliminvisible
\isataginvisible
\isacommand{termination}\isamarkupfalse%
\isanewline
\ \ \isacommand{apply}\isamarkupfalse%
\ {\isacharparenleft}{\kern0pt}relation\ {\isacartoucheopen}tak{\isacharunderscore}{\kern0pt}m{\isadigit{1}}\ {\isacharless}{\kern0pt}{\isacharasterisk}{\kern0pt}mlex{\isacharasterisk}{\kern0pt}{\isachargreater}{\kern0pt}\ tak{\isacharunderscore}{\kern0pt}m{\isadigit{2}}\ {\isacharless}{\kern0pt}{\isacharasterisk}{\kern0pt}mlex{\isacharasterisk}{\kern0pt}{\isachargreater}{\kern0pt}\ tak{\isacharunderscore}{\kern0pt}m{\isadigit{3}}\ {\isacharless}{\kern0pt}{\isacharasterisk}{\kern0pt}mlex{\isacharasterisk}{\kern0pt}{\isachargreater}{\kern0pt}\ {\isacharbraceleft}{\kern0pt}{\isacharbraceright}{\kern0pt}{\isacartoucheclose}{\isacharparenright}{\kern0pt}\ \isanewline
\ \ \ \ \ \ \isacommand{apply}\isamarkupfalse%
\ {\isacharparenleft}{\kern0pt}simp\ add{\isacharcolon}{\kern0pt}\ wf{\isacharunderscore}{\kern0pt}mlex{\isacharparenright}{\kern0pt}\ \isanewline
\ \ \ \ \ \isacommand{apply}\isamarkupfalse%
\ {\isacharparenleft}{\kern0pt}simp\ add{\isacharcolon}{\kern0pt}\ l{\isacharunderscore}{\kern0pt}call{\isadigit{1}}{\isacharparenright}{\kern0pt}\isanewline
\ \ \ \ \ \isacommand{apply}\isamarkupfalse%
\ {\isacharparenleft}{\kern0pt}simp\ add{\isacharcolon}{\kern0pt}\ l{\isacharunderscore}{\kern0pt}call{\isadigit{2}}{\isacharparenright}{\kern0pt}\isanewline
\ \ \ \ \ \isacommand{apply}\isamarkupfalse%
\ {\isacharparenleft}{\kern0pt}simp\ add{\isacharcolon}{\kern0pt}\ l{\isacharunderscore}{\kern0pt}call{\isadigit{3}}{\isacharparenright}{\kern0pt}\isanewline
\ \ \isacommand{by}\isamarkupfalse%
\ {\isacharparenleft}{\kern0pt}simp\ add{\isacharcolon}{\kern0pt}\ l{\isacharunderscore}{\kern0pt}call{\isadigit{4}}{\isacharparenright}{\kern0pt}%
\endisataginvisible
{\isafoldinvisible}%
\isadeliminvisible
\endisadeliminvisible
\begin{isamarkuptext}%
With the total version of induction and simplification rules are
available. Then, it is possible to prove its rather simpler equivalence.%
\end{isamarkuptext}\isamarkuptrue%
\isacommand{theorem}\isamarkupfalse%
\ tak{\isacharunderscore}{\kern0pt}correct{\isacharcolon}{\kern0pt}\ {\isacartoucheopen}tak\ x\ y\ z\ {\isacharequal}{\kern0pt}\ {\isacharparenleft}{\kern0pt}if\ x\ {\isasymle}\ y\ then\ y\ else\ if\ y\ {\isasymle}\ z\ then\ z\ else\ x{\isacharparenright}{\kern0pt}{\isacartoucheclose}\isanewline
\isadelimproof
\ \ %
\endisadelimproof
\isatagproof
\isacommand{by}\isamarkupfalse%
\ {\isacharparenleft}{\kern0pt}induction\ x\ y\ z\ rule{\isacharcolon}{\kern0pt}\ tak{\isachardot}{\kern0pt}induct{\isacharparenright}{\kern0pt}\ auto%
\endisatagproof
{\isafoldproof}%
\isadelimproof
\endisadelimproof
\isadelimdocument
\endisadelimdocument
\isatagdocument
\isamarkupsubsection{Mutual recursion\label{subsec:MutualRec}%
}
\isamarkuptrue%
\endisatagdocument
{\isafolddocument}%
\isadelimdocument
\endisadelimdocument
\begin{isamarkuptext}%
Finally, we handle mutual recursion. VDM has few bounds on mutually
recursive definitions. We reuse the VDM type checker cyclic-measure discovery
algorithm in order to deduce related recursive calls. Note this is not
complete and may fail to discover involved recursive calls, such as those
under some lambda-term result, for example. If that recursive call graph
discovery fails, then users have to annotate at least one of the mutually
recursive definitions with an \isatt{@{\kern0pt}I{\kern0pt}s{\kern0pt}a{\kern0pt}M{\kern0pt}u{\kern0pt}t{\kern0pt}u{\kern0pt}a{\kern0pt}l{\kern0pt}R{\kern0pt}e{\kern0pt}c{\kern0pt}} annotation listing all expected
function names involved.

Mutually recursive definitions proof obligations refer to each others
measure functions. Isabelle requires all related functions to be given in a
single definition, where Isabelle's sum (or union) types are used in the proof
setup. Where needed, termination proof setup will be involved and with limited 
automation. Here is a simple example and its translation, which works with
 \isa{fun} syntax, hence has no further proof obligation needs. Note that in 
this case using \isa{fun} syntax, we declare implicit preconditions as 
simplification rules.
\begin{vdmsl}[frame=none,basicstyle=\ttfamily\scriptsize]
    --@IsaMutualRec({odd})
    even: nat -> bool
    even(n) == if n = 0 then true else odd(n-1) measure n;
    
    --@IsaMutualRec({even})
    odd: nat -> bool
    odd(n) == if n = 0 then false else even(n-1) measure n;
\end{vdmsl}%
\end{isamarkuptext}\isamarkuptrue%
\isacommand{definition}\isamarkupfalse%
\ pre{\isacharunderscore}{\kern0pt}even\ {\isacharcolon}{\kern0pt}{\isacharcolon}{\kern0pt}\ {\isacartoucheopen}VDMNat\ {\isasymRightarrow}\ {\isasymbool}{\isacartoucheclose}\ \isakeyword{where}\ {\isacharbrackleft}{\kern0pt}termination{\isacharunderscore}{\kern0pt}simp{\isacharbrackright}{\kern0pt}{\isacharcolon}{\kern0pt}\ {\isacartoucheopen}pre{\isacharunderscore}{\kern0pt}even\ n\ {\isasymequiv}\ inv{\isacharunderscore}{\kern0pt}VDMNat\ n{\isacartoucheclose}\ \isanewline
\isacommand{definition}\isamarkupfalse%
\ pre{\isacharunderscore}{\kern0pt}odd\ \ {\isacharcolon}{\kern0pt}{\isacharcolon}{\kern0pt}\ {\isacartoucheopen}VDMNat\ {\isasymRightarrow}\ {\isasymbool}{\isacartoucheclose}\ \isakeyword{where}\ {\isacharbrackleft}{\kern0pt}termination{\isacharunderscore}{\kern0pt}simp{\isacharbrackright}{\kern0pt}{\isacharcolon}{\kern0pt}\ {\isacartoucheopen}pre{\isacharunderscore}{\kern0pt}odd\ n\ {\isasymequiv}\ inv{\isacharunderscore}{\kern0pt}VDMNat\ n{\isacartoucheclose}\ \isanewline
\isacommand{fun}\isamarkupfalse%
\ {\isacharparenleft}{\kern0pt}domintros{\isacharparenright}{\kern0pt}\ even\ {\isacharcolon}{\kern0pt}{\isacharcolon}{\kern0pt}\ {\isacartoucheopen}VDMNat\ {\isasymRightarrow}\ {\isasymbool}{\isacartoucheclose}\ \isakeyword{and}\ odd\ \ {\isacharcolon}{\kern0pt}{\isacharcolon}{\kern0pt}\ {\isacartoucheopen}VDMNat\ {\isasymRightarrow}\ {\isasymbool}{\isacartoucheclose}\ \isakeyword{where}\ \isanewline
\ \ {\isacartoucheopen}even\ n\ {\isacharequal}{\kern0pt}\ {\isacharparenleft}{\kern0pt}if\ pre{\isacharunderscore}{\kern0pt}even\ n\ then\ {\isacharparenleft}{\kern0pt}if\ n\ {\isacharequal}{\kern0pt}\ {\isadigit{0}}\ then\ True\ else\ odd\ {\isacharparenleft}{\kern0pt}n{\isacharminus}{\kern0pt}{\isadigit{1}}{\isacharparenright}{\kern0pt}{\isacharparenright}{\kern0pt}\ else\ undefined{\isacharparenright}{\kern0pt}{\isacartoucheclose}\isanewline
{\isacharbar}{\kern0pt}\ {\isacartoucheopen}odd\ \ n\ {\isacharequal}{\kern0pt}\ {\isacharparenleft}{\kern0pt}if\ pre{\isacharunderscore}{\kern0pt}odd\ n\ then\ {\isacharparenleft}{\kern0pt}if\ n\ {\isacharequal}{\kern0pt}\ {\isadigit{0}}\ then\ False\ else\ even\ {\isacharparenleft}{\kern0pt}n{\isacharminus}{\kern0pt}{\isadigit{1}}{\isacharparenright}{\kern0pt}{\isacharparenright}{\kern0pt}\ else\ undefined{\isacharparenright}{\kern0pt}{\isacartoucheclose}%
\isadelimdocument
\endisadelimdocument
\isatagdocument
\isamarkupsection{Discussion and conclusion\label{sec:Conclusion}%
}
\isamarkuptrue%
\endisatagdocument
{\isafolddocument}%
\isadelimdocument
\endisadelimdocument
\begin{isamarkuptext}%
This paper has presented a translation strategy from VDM to Isabelle for
recursive functions over basic types, sets and maps, as well as mutual
recursion. We present how the strategy works for complex recursion.

The complex recursion examples hint at possible VDM recursive measure
extensions to use a combination of measure relations and functions. The full
VDM and Isabelle sources and proofs can be found at the VDM toolkit repository
at \isatt{R{\kern0pt}e{\kern0pt}c{\kern0pt}u{\kern0pt}r{\kern0pt}s{\kern0pt}i{\kern0pt}v{\kern0pt}e{\kern0pt}V{\kern0pt}D{\kern0pt}M{\kern0pt}*{\kern0pt}.{\kern0pt}t{\kern0pt}h{\kern0pt}y{\kern0pt}}\footnote{\url{https://github.com/leouk/VDM_Toolkit}}.

\textbf{Future work.}~We are implementing the translation strategy in the
\isatt{v{\kern0pt}d{\kern0pt}m{\kern0pt}2{\kern0pt}i{\kern0pt}s{\kern0pt}a{\kern0pt}} plugin, which should be available soon.

\textbf{Acknowledgements.}~We appreciated discussions with Stephan Merz on pointers
for complex well-founded recursion proofs in Isabelle, and with Nick Battle on
limits for VDM recursive measures. Peter Gorm Larsen would also like to
acknowledge the Poul Due Jensen Foundation for funding the project Digital
Twins for Cyber-Physical Systems (DiT4CPS).%
\end{isamarkuptext}\isamarkuptrue%
\isadeliminvisible
\endisadeliminvisible
\isataginvisible
\endisataginvisible
{\isafoldinvisible}%
\isadeliminvisible
\endisadeliminvisible
\isadelimtheory
\endisadelimtheory
\isatagtheory
\endisatagtheory
{\isafoldtheory}%
\isadelimtheory
\endisadelimtheory
\end{isabellebody}%

\bibliographystyle{splncs03}
 \bibliography{root}

\begin{thebibliography}{1}
\providecommand{\url}[1]{\texttt{#1}}
\providecommand{\urlprefix}{URL }

\bibitem{Battle09}
Battle, N.: {VDMJ User Guide}. Tech. rep., Fujitsu Services Ltd., UK (2009)

\bibitem{VDMJAnnotations}
Battle, N.: VDMJ Annotaions Guide (2022)

\bibitem{SCNP_POPL}
Ben-Amram, A.M., Codish, M.: A sat-based approach to size change termination
  with global ranking functions. In: TACAS. pp. 218--232. Springer (2008)

\bibitem{NimFull}
Freitas, L.: A {VDM} translation strategy to isabelle ({Nill Full} game) (2016)

\bibitem{IsaFunctionPackage}
Krauss, A.: Defining Recursive Functions in Isabelle/HOL. Technical University
  of Munich (2021)

\bibitem{KrausSCNP}
Krauss, A., Heller, A.: A mechanized proof reconstruction for scnp termination.
  Tech. rep., Technical University of Munich (2012)

\bibitem{SCT_POPL}
Lee, C.S., Jones, N.D., Ben-Amram, A.M.: The size-change principle for program
  termination. In: POPL. pp. 81--92. POPL '01, ACM, New York USA (2001)

\bibitem{AdvancedVSCodePaper}
Rask, J.K., et~al.: {Advanced VDM Support in Visual Studio Code}. In: 20th
  International Overture Workshop. pp. 35--50 (2022)

\end{thebibliography}

\end{document}